\documentclass[sn-nature, linenumbers, 12pt]{article}
\usepackage{amsfonts}
\usepackage{amssymb}
\usepackage{amsmath}
\usepackage{booktabs}
\usepackage{graphicx}
\usepackage{dcolumn}
\usepackage{bm}
\usepackage{units}
\usepackage{multirow}
\usepackage{CJKutf8}
\usepackage{subfigure}
\usepackage{color}
\usepackage{xcolor}
\usepackage{url}
\usepackage{geometry}
\usepackage{setspace}
\usepackage[utf8]{inputenc}
\usepackage[style = nature, backend=biber, autocite=superscript, doi = false, url = false, date=year, isbn=false = false]{biblatex}
\usepackage[colorlinks,linkcolor=blue,anchorcolor=blue,citecolor=blue,urlcolor=blue]{hyperref}

\bibliography{uni-jde-plainbib-V2}

\usepackage{authblk}
\author[1]{Haowei Ye\textsuperscript{\textdagger}}
\author[1]{Wenxue He\textsuperscript{\textdagger}}
\author[1]{Kaixuan Fan}
\author[1]{Yingpeng Zhang}
\author[1]{Shijin Li}
\author[1]{Yu Pan}
\author[2,3,4]{Dechao Geng}
\author[1]{Fan Yang}
\author[5]{Kenji Watanabe}
\author[6]{Takashi Taniguchi}
\author[1,7]{Hechen Ren *}
\affil[1]{Center for Joint Quantum Studies \& Tianjin Key Laboratory of Low Dimensional Materials Physics and Preparing Technology, Department of Physics, School of Science, Tianjin University, Tianjin 300072, China}
\affil[2]{State Key Laboratory of Advanced Materials for Intelligent Sensing, Ministry of Science and Technology \& Key Laboratory of Organic Integrated Circuit, Ministry of Education \& Tianjin Key Laboratory of Molecular Optoelectronic Sciences, Department of Chemistry, School of Science, Tianjin University, Tianjin 300072, China}
\affil[3]{Collaborative Innovation Center of Chemical Science and Engineering, Tianjin 300072, China}
\affil[4]{Haihe Laboratory of Sustainable Chemical Transformations, Tianjin 300192, China}
\affil[5]{Research Center for Electronic and Optical Materials, National Institute for Materials Science, Tsukuba, Japan}
\affil[6]{Research Center for Materials Nanoarchitectonics, National Institute for Materials Science, Tsukuba, Japan}
\affil[7]{Joint School of National University of Singapore and Tianjin University, International Campus of Tianjin University, Binhai New City, Fuzhou 350207, China}

\title{Magnetic-Free Quantum Interference and Universal Josephson Diode Effect Driven by a Supercurrent Gauge Field}
\date{\textsuperscript{\textdagger} These authors contributed equally to this work. \\ * e-mail: ren@tju.edu.cn}

\begin{document}

\maketitle

\begin{spacing}{1.5} 
 
\newpage
\begin{abstract}
The Josephson effect, a hallmark of superconducting phase coherence, drives modern quantum technologies. However, Josephson-based quantum interference has hitherto been tethered to magnetic fields, despite phase coherence being a quintessential, intrinsic trait of superconductivity. Moreover, the Josephson diode effect (JDE) is typically viewed as an anomalous phenomenon indicative of broken symmetries in exotic phases of matter. Here, in planar Josephson junctions made with $\mathrm{Bi}_2\mathrm{O}_2\mathrm{Se}$ and bilayer graphene, we demonstrate that the JDE is a missing universal property of the Josephson effect. Simultaneously, we present an all-electric technology that replaces magnetic flux for controlling and measuring supercurrent interference. Central to our approach is a supercurrent gauge field (SGF), generated and amplified through high-kinetic-inductance superconductors and novel device architectures. By establishing the physical equivalence between the SGF and a magnetic field, we eliminate the reliance on external fields in quantum interference and reveal a universal, field-free JDE mechanism with broad implications for detecting broken-symmetry states. Finally, we show that the SGF offers capabilities beyond those of a conventional magnetic field by experimentally demonstrating a magnetic-free, phase-sensitive technique to construct and characterize finite-momentum superconductivity, opening new frontiers for exploring novel phases of matter and superconducting quantum architectures.
\end{abstract}

\newpage
\par

The Josephson effect remains a cornerstone of macroscopic quantum phenomena and a primary driver of modern quantum technologies \autocite{Martinis1985, Devoret1985}, as highlighted by the 2025 Nobel Prize in Physics for leveraging superconducting phase coherence in metrology and quantum computing \autocite{Dobrovolskiy2026}. In the presence of a magnetic field, the critical current of Josephson junctions and SQUIDs exhibits periodic oscillations in magnetic flux \autocite{Tinkham2004}. However, this textbook paradigm of quantum interference conflates the phase coherence innate to superconductivity with the necessity of an external magnetic field. To this day, an overwhelming reliance on magnetic flux persists across almost all Josephson-based architectures, including tunable transmons, flux qubits, and fluxonium qubits \autocite{Fu2026, Rower2023, Nguyen2019}, introducing magnetic noise and engineering complexity. Furthermore, proposals for realizing exotic phases of matter await a scalable platform for arbitrary phase generation. For example, breaking time-reversal symmetry solely via superconducting phase bias can reliably induce topological superconductivity \autocite{Melo2019, Lesser2021, Lesser2024}, removing the Zeeman field required in established setups \autocite{Ren2019, Loo2026}. Yet, these proposals still rely on the Aharonov-Bohm effect in flux-biased rings, again inducing magnetic crosstalk during scaling. Therefore, decoupling quantum interference from magnetic fields represents both a vital paradigm shift in fundamental physics and a transformative leap for magnetic-free quantum technologies.

Meanwhile, the Josephson diode effect represents a shift from symmetric macroscopic tunneling to nonreciprocal quantum transport, enabled by the simultaneous breaking of time-reversal (TR) and inversion symmetries \autocite{Ando2020, Wakatsuki2017, Yuan2022}. Examples include noncentrosymmetric InAs quantum wells \autocite{Baumgartner2021} and type-II Dirac and Weyl semimetals \autocite{Pal2022, Sivakumar2024, Kim2024}, where an external magnetic field breaks TR symmetry. Recent efforts have pivoted toward field-free JDE, exploring mechanisms such as trapped flux and asymmetric geometries \autocite{Golod2022, Chen2024d, Hou2023, Guarcello2024}, ferromagnetic barriers \autocite{Jeon2022, Wu2022a, Trahms2023}, or spontaneous symmetry breaking in exotic materials \autocite{DiezMerida2023, Ma2025a}. Another approach involves Andreev molecules through multi-terminal junctions \autocite{Gupta2023a, Matsuo2023a, Chiles2023, Zhu2025}. Despite these advances, a universal mechanism for the JDE remains missing from the standard Josephson model, hindering the development of magnetic-free, scalable superconducting diodes.

Here, we tackle both challenges with a unified approach. By engineering superconducting leads for high-density phase gradients, we observe nonreciprocal switching currents in planar Josephson junctions fabricated from centrosymmetric $\mathrm{Bi}_2\mathrm{O}_2\mathrm{Se}$ and bilayer graphene. The diode polarity reverses with the sign of the flux, yielding a skewed Fraunhofer envelope in the supercurrent interference pattern (SIP). We attribute this to a supercurrent gauge field that drives a phase-winding effect analogous to a physical magnetic field, breaking time-reversal symmetry for a universal diode response. By demonstrating this effect in a standard device geometry, we redefine the Josephson junction as an intrinsic diode, offering a material-agnostic blueprint for supercurrent rectification. Furthermore, in a novel junction designed to amplify the SGF, we reproduce both symmetric and skewed Fraunhofer patterns using $\mu$A-scale electrical currents, achieving macroscopic quantum interference and the JDE entirely free of magnetic fields. Finally, we leverage the SGF's versatility to induce and quantitatively probe finite-momentum superconductivity. Our innovative, phase-sensitive technology opens new frontiers for exotic phases of matter and magnetic-free quantum architectures.

\section*{Universal Josephson diode}

Although Josephson junctions fabricated from centrosymmetric $\mathrm{Bi}_2\mathrm{O}_2\mathrm{Se}$ and hexagonal boron nitride (hBN)-encapsulated bilayer graphene have been previously reported \autocite{Kraft2018, Rout2024, Ying2020}, the Josephson diode effect has not been observed in these time-reversal-symmetric systems. Here, we contact these materials using Al-Au bilayers, which possess high kinetic inductance and are widely applicable in microwave technologies such as photon detection \autocite{Hu2020, Wang2023c}. To mimic the inductive boost of granular aluminum \autocite{Valenti2019} without introducing oxygen, we dope the Al layer with Cu. Each superconducting lead extends on both sides to two separate wire connections, forming a standard four-terminal geometry. We begin by presenting transport data from our $\mathrm{Bi}_2\mathrm{O}_2\mathrm{Se}$ junctions (Fig. 1a).

There are two configurations to source current in a four-terminal Josephson junction: input current and ground on diagonally opposite corners (Fig. 1a) or on the same side (Fig. 1b,c). In both schemes, the voltage across the junction is measured using the remaining two wires. Throughout the literature, these configurations have routinely been treated as equivalent under the assumption that the two wires connected to a single lead are shorted by the superconductor. However, if we move beyond treating the phase of each lead as a uniform, single variable and instead regard it as a spatially varying local gauge, these two configurations differ fundamentally in their symmetry. Hereafter, we refer to the diagonal arrangement as the symmetric configuration and the same-side arrangement as the asymmetric configuration.

\begin{figure*}[htbp]
\centering
\includegraphics[width=0.9\linewidth]{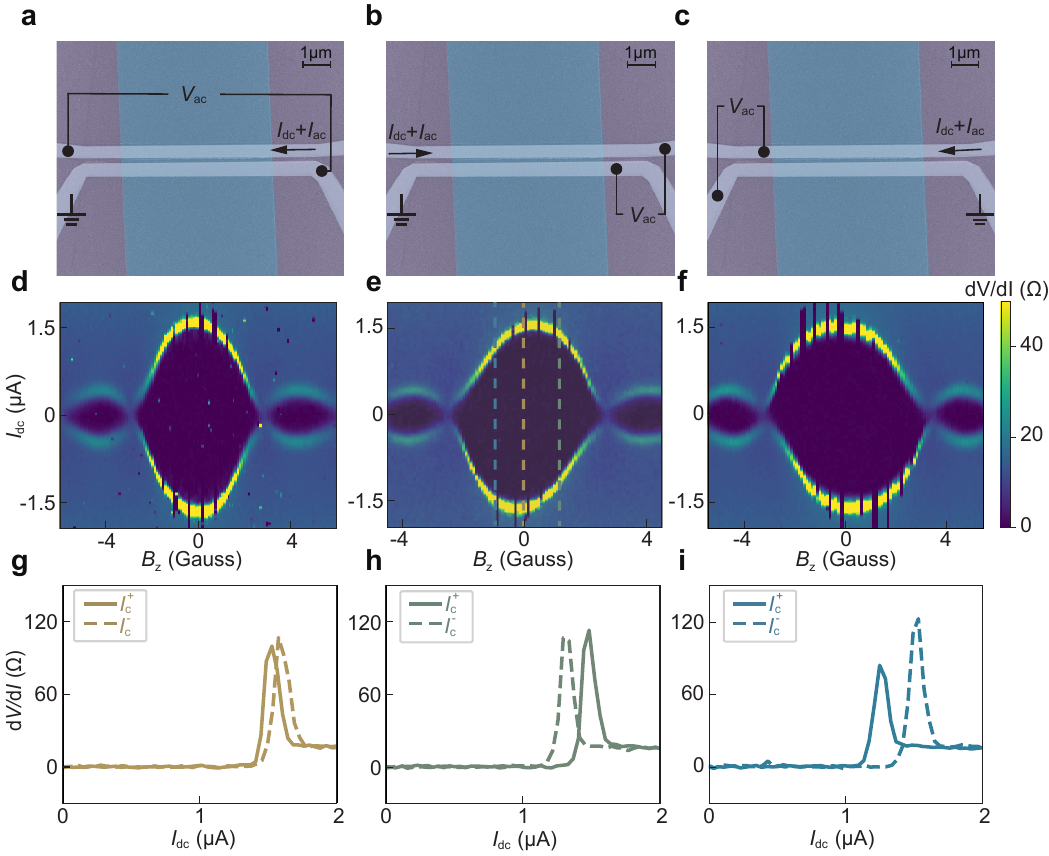} 
\caption{\textbf{Universal Josephson diode effect (JDE) in a $\mathrm{Bi}_2\mathrm{O}_2\mathrm{Se}$ Josephson junction (JJ1).}
\textbf{a–c}, False-colour scanning electron microscope (SEM) images of JJ1 showing three distinct current-flow configurations: symmetric \textbf{a} and asymmetric \textbf{b–c}.
\textbf{d}, Normal Fraunhofer interference pattern measured in the symmetric configuration \textbf{a}, exhibiting no JDE.
\textbf{e, f}, Anomalous Fraunhofer patterns with pronounced JDE obtained using the asymmetric configurations \textbf{b} and \textbf{c}, respectively.
\textbf{g–i}, Linecuts of differential resistance $dV/dI$ vs. bias current $I_{dc}$ extracted from \textbf{e} at $B_z = 0$~G, $1.1$~G, and $-0.9$~G, illustrating the field-induced polarity reversal of the JDE. $I^+$ and $I^-$ indicate positive and negative sweeping direcitons.}
\label{fig_1}
\end{figure*}

In the symmetric configuration, the measured switching current features a conventional Fraunhofer pattern in an out-of-plane magnetic field. At all field values, the positive and negative switching currents ($I_{\mathrm{c}}^+$, $I_{\mathrm{c}}^-$) are equal in magnitude, showing no JDE (Fig. 1d). Conversely, in the asymmetric configuration, we observe a skewed interference pattern with lifted nodes (Fig. 1e,f). The maximum of $I_{\mathrm{c}}^+$ is shifted in magnetic field relative to that of $I_{\mathrm{c}}^-$. Consequently, while the transport remains reciprocal at precisely zero flux (Fig. 1g), a pronounced JDE emerges across a wide range of non-zero magnetic fields. In positive fields, $I_{\mathrm{c}}^+$ exceeds the magnitude of $I_{\mathrm{c}}^-$ (Fig. 1h), whereas $I_{\mathrm{c}}^-$ prevails in negative fields (Fig. 1i). This JDE persists across a broad range of temperatures and in-plane magnetic fields, maintaining a distinct odd parity with respect to the magnetic flux (Extended Data Figs. S1-S3).

At first glance, this behavior might be attributed to a self-field effect arising from the asymmetric current configuration. However, the calculated self-field generated by a $2 \; \mathrm{\mu}$A current in the junction is on the order of $2.5 \; \mathrm{\mu}$T---nearly two orders of magnitude smaller than the applied external field. It is therefore highly unlikely that the self-field could shift the critical current maxima by a significant fraction of a flux quantum, which easily corresponds to $2.5$~G. Although the induced magnetic field from the current is negligible, a more fundamental mechanism tied to the macroscopic quantum properties of the superconducting order parameter drives this behavior, which we outline below.

\section*{Magnetic-free JDE and quantum interference}

In Ginzburg-Landau theory, the complex Cooper-pair wavefunction $\psi = |\psi| e^{i\phi}$ minimizes the free energy of the superconductor in the presence of fields, currents, and gradients. According to the first Ginzburg-Landau equation, the local supercurrent density is given by

\begin{equation} 
\mathbf{J}_s = \frac{q}{m^*} |\psi|^2 \left( \hbar \mathbf{\nabla} \phi - \frac{q}{c}\mathbf{A} \right) ,
\end{equation}
where $q = 2e$ is the electrical charge of the Cooper pair, $m^* = 2 m_e$ its effective mass, $\hbar$ is the reduced Planck constant, $c$ is the speed of light, and $\mathbf{A}$ is the vector potential of the external magnetic field \autocite{Tinkham2004}. 

Because the supercurrent density is a physical observable, the expression within the parentheses must remain gauge-invariant. This signifies that the gauge symmetry in the magnetic field is now shared with the superconductor's order parameter and both the vector potential and the local supercurrent dictate the spatial phase gradient. This becomes more transparent by rearranging the terms to isolate the phase gradient:

\begin{equation} 
\mathbf{\nabla} \phi = \frac{q}{\hbar c}\mathbf{A} + \frac{m^*}{\hbar q n_s} \mathbf{J}_s ,
\end{equation}
where $n_s = |\psi|^2$ is the Cooper-pair density. By defining the vector potential for the SGF as $\mathbf{A}_s = \frac{m^*c}{n_s q^2}\mathbf{J}_s$, the combined gauge field $\mathbf{A} + \mathbf{A}_s$ captures the total phase-driving potential of the system. This formulation explicitly underscores the equivalence between an external magnetic field and a supercurrent which allows them to cooperate, interfere, and replace each other.

\begin{figure*}[htbp]
\centering
\includegraphics[width=0.9\linewidth]{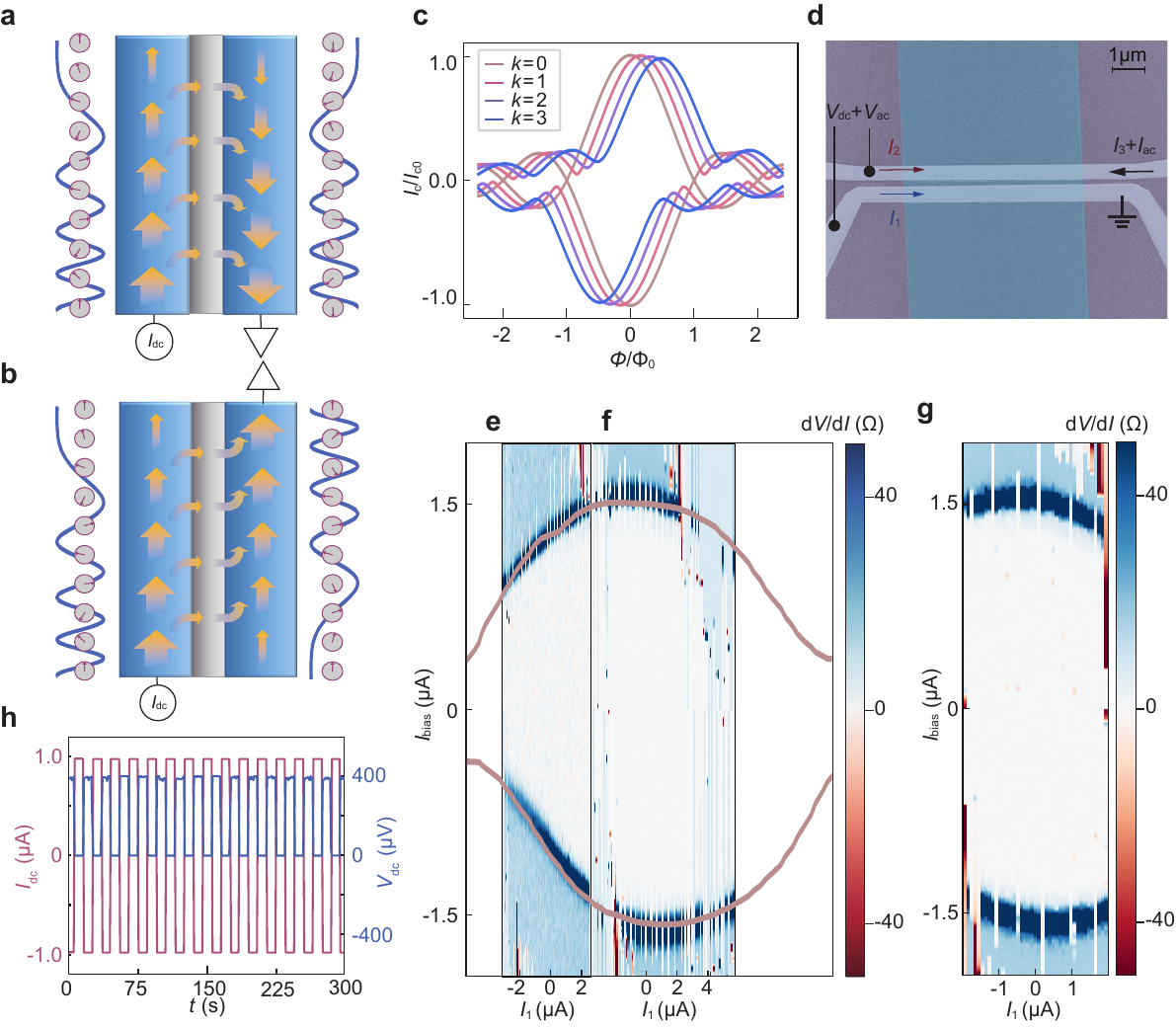} 
\caption{\textbf{Illustration of supercurrent gauge fields (SGF) and magnetic-free quantum interference.}
\textbf{a, b}, Schematics illustrating the current flows and corresponding SGFs in asymmetric \textbf{a} and symmetric \textbf{b} configurations.
\textbf{c}, Calculated evolution of $I_c(B_z)$ with increasing SGF under asymmetric current injection, showing the lateral shift of the SIP.
\textbf{d}, Schematic of the measurement configuration for JJ1 with control currents $I_{1}$ and $I_{2}$, and DC bias $I_{\text{bias}} = I_{3} + I_{2}$.
\textbf{e, f}, SIPs of $dV/dI$ vs. $I_{\mathrm{bias}}$ and control $I_{1}$ (with $I_{2} = 0$) at $B_z = -1.5$~G \textbf{e} and $0$~G \textbf{f}, demonstrating SGF-induced modulation of the interference pattern. $I_c(B_z)$ envelope (pale pink) is extracted from the data in Fig. 1f.
\textbf{g}, Zero-field SIP obtained with two-terminal control ($I_{1} = -I_{2}$) and bias $I_{\text{bias}} = I_{3} + I_{2}$, demonstrating fully magnetic-field-free quantum interference and JDE.
\textbf{h}, Measured voltage drop (blue) under square-wave current pulse (pink) at zero magnetic field, with $I_{1} = -2\ \mu\text{A}$ and $I_{2} = 2\ \mu\text{A}$, showing robust rectifying behaviour at zero field.}
\label{fig_2}
\end{figure*}

Having laid the foundation, we reexamine the supercurrent interference effect in a standard Josephson junction. Adopting the Landau gauge $\mathbf{A} = (0, B_z x, 0)$, the phase difference across the weak link at each $x$-position is

\begin{equation} 
\gamma(x) = \phi_2(x) - \phi_1(x) + \frac{2 \pi d}{\Phi_0} B_z x ,
\label{phase_diff_x}
\end{equation}
where $d$ is the junction length and $\Phi_0 = \frac{h}{2e}$ is the magnetic flux quantum \autocite{Hart2014}. Along the superconducting leads, the spatial derivatives of the phases are governed by the local currents:

\begin{equation} 
\frac{\partial \phi_i(x)}{\partial x} = L_k I_i(x), \; i = 1,2,
\label{lead_derivative}
\end{equation} 
where $L_k = \frac{m^*}{\hbar q S n_s} $ represents the kinetic inductance per unit length of the leads, with $S$ being their cross-sectional area. Assuming for simplicity that the currents $I_1$ and $I_2$ remain constant along the leads, we can define an effective wavenumber $k_{\mathrm{tot}}$ that captures the combined phase-winding forces:

\begin{equation} 
k_{\mathrm{tot}} = L_k (I_2 - I_1) + k_B,
\end{equation}
where $k_B = \frac{2 \pi d}{\Phi_0} B_z$. The total critical current $I_{\mathrm{c}}$ is obtained by integrating the local Josephson current density over the junction width $W$ and maximizing with respect to the global phase difference $\gamma_0$:

\begin{equation} 
I_c =  \max_{\gamma_0} \int_{-W/2}^{W/2} J_c \sin(\gamma_0 + k_{\mathrm{tot}}x) \mathrm{d}x = I_{c0} \left| \mathrm{sinc} \left( \frac{k_{\mathrm{tot}} W}{2} \right) \right|,
\label{ic_int}
\end{equation}

where $J_c$ is the critical current density and $I_{\mathrm{c0}}$ is the junction's maximum critical current. In the symmetric configuration, $I_2$ and $I_1$ flow in the same direction with equal magnitude ($I_1 = I_2 = I_{\mathrm{bias}}$), so $k_{\mathrm{tot}} = k_B$, recovering the conventional Fraunhofer pattern. In the asymmetric configuration, the currents flow in opposite directions ($I_2 = -I_1 = I_{\mathrm{bias}}$), which leads to a current-dependent total wavenumber $k_{\mathrm{tot}} = 2 L_k I_{\mathrm{bias}} + k_B$. Under the same $k_B$, positive and negative bias currents shift $k_{\mathrm{tot}}$ in opposite directions. This produces a lateral splitting in the magnetic field dependence of the positive and negative branches of the SIP, generating the observed diode effect.

In a four-terminal Josephson junction, as the supercurrent propagates along each lead, it continuously enters or exits the lead to traverse the junction (Fig. 2a,b). To first order, this results in a linear gradient of supercurrent density along each lead. This introduces a quadratic term alongside the linear term in the phase profiles $\phi_i(x)$, effectively transforming the conventional Fraunhofer diffraction pattern into a Fresnel diffraction pattern. Numerical simulations reveal partially lifted nodes in the resulting SIP alongside the lateral shift between the positive and negative branches (Fig. 2c). These features are visible in our experimental data (Fig. 1e).

To verify this theoretical framework, we apply an additional control current, $I_{\mathrm{ctrl}}$, along the grounded superconducting lead by connecting a current source to its opposite end (Fig. 2d). This creates a uniform SGF and thereby a linear phase winding along the grounded lead. This adds to the effect of the magnetic field directly by modifying the total wavenumber to $k_{\mathrm{tot}} = L_k (I_2 - I_1) + L_k I_{\mathrm{ctrl}} + k_B$. At any fixed external flux, varying $I_{\mathrm{ctrl}}$ modulates the critical currents in a manner equivalent to sweeping the external magnetic flux (Fig. 2e, f), reproducing segments of the original SIP without actually tuning the magnetic field. By comparing the slopes of the field-modulated and control-current-modulated critical currents, we extract a lead kinetic inductance of $0.0786\ \mu\mathrm{m}^{-1}\,\mu\mathrm{A}^{-1}$. Given the leads' cross-sectional area is approximately $3.15 \times 10^{4}\ \text{nm}^{2}$, this value is over $20$ times the expected inductance of pure aluminum \autocite{Valenti2019}. We attribute this inductive boost to a combination of the Cu-induced granularity of Al and a reverse proximity effect from the Au layer and the intermetallic region, both diluting the Cooper-pair density $n_s$ in the superconducting leads and reducing their superfluid stiffness. This interpretation is evidenced by the breakdown condition of the combined bias and control currents, which occurs just under $4 \; \mu$A. To amplify this effect, we can apply counter-propagating control currents to both leads, thereby doubling the SGF contribution. Under this configuration, the critical current modulation becomes more pronounced (Fig. 2g), achieving a greater zero-field JDE within the same range. Figure 2h illustrates on-off cycles of this field-free diode, showing repeatable rectifying behavior.

\begin{figure*}[htbp]
\centering
\includegraphics[width=0.9\textwidth]{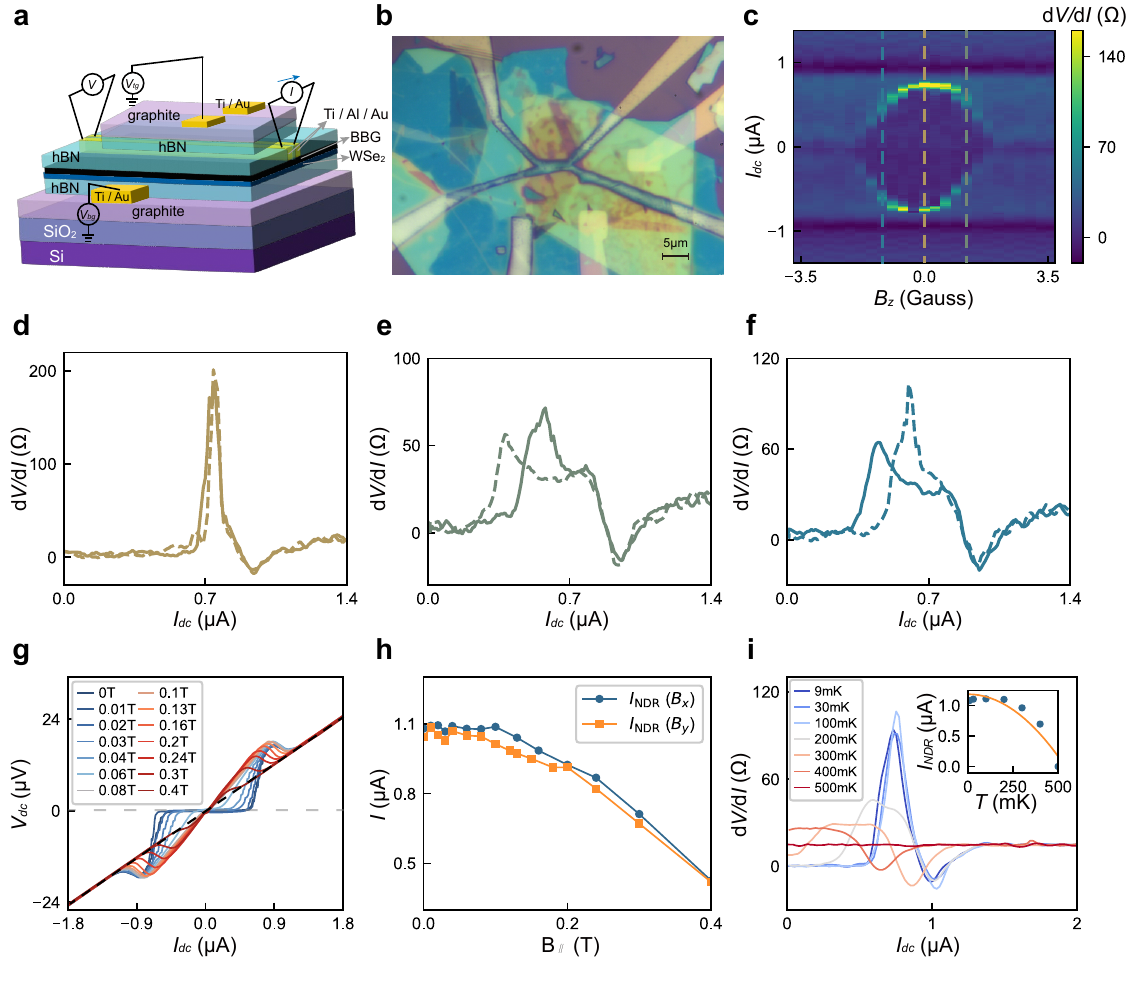} 
\caption{\textbf{Josephson diode effect (JDE) in a Bernal bilayer graphene (BBG) Josephson junction.}
\textbf{a}, Schematic of the van der Waals heterostructure stack.
\textbf{b}, Optical micrograph of the BBG Josephson junction device.
\textbf{c}, SIP of differential resistance $dV/dI$ vs. perpendicular field $B_z$ and bias current $I_{\mathrm{dc}}$ at $V_{\text{bg}} = 3$~V and $V_{\text{tg}} = 0$~V. Beige, olive, and teal dashed lines mark $B_z = 0$, $1.2$, and $-1.2$~G. \textbf{d--f}, Corresponding line cuts taken from \textbf{c} at $B_z = 0$~G \textbf{d}, $1.2$~G \textbf{e} and $-1.2$~G \textbf{f}, displaying $dV/dI$ vs. $I_{\mathrm{dc}}$. Solid and dashed curves represent $I^+$ and $I^-$, respectively. At $B_z = 0$~G, no JDE is observed \textbf{d}. At $B_z = 1.2$~G, a JDE appears with $I_{\mathrm{c}}^+ > I_{\mathrm{c}}^-$ \textbf{e}. At $B_z = -1.2$~G, the JDE appears with opposite polarity: $I_{\mathrm{c}}^+ < I_{\mathrm{c}}^-$ \textbf{f}.
\textbf{g}, Current–voltage ($I\text{-}V$) characteristics vs. in-plane field $B_x$ (parallel to current direction) at $V_{\text{tg}} = 0$~V and $V_{\text{bg}} = 2$~V, showing the gradual suppression of the NDR region with increasing $B_x$.
\textbf{h}, Extracted NDR width $I_{\text{NDR}}$ vs. $B_x$ (from panel \textbf{g}) and $B_y$ (perpendicular to current direction, see Extended Data Fig. S4).
\textbf{i}, $dV/dI$ vs. $I_{\mathrm{dc}}$ at various temperatures, showing gradual decrease of the NDR with increasing temperature. Inset: Temperature dependence of $I_{\text{NDR}}$ (blue dots) fitted with $I_{\text{NDR}}(T) = I_{\text{NDR}}(0)[1-(T/T_c)^2]$ (orange curve), yielding $I_{\text{NDR}}(0) = 1.187 \pm 0.071\ \mu\text{A}$ and $T_c = 539.3 \pm 31.5$~mK ($R^2 = 0.9015$).}
\label{fig_3}
\end{figure*}

The flux-odd JDE phenomenon is reproduced in our bilayer graphene Josephson junction (Fig. 3a,b), where the SIP exhibits a familiar skewed profile (Fig. 3c). Again, the positive and negative critical currents remain equal at zero flux and diverge at finite fields, showing opposite polarities under opposite fluxes (Fig. 3d-f). Another intriguing feature is the appearance of negative differential resistance (NDR) once the bias current surpasses the critical current on both the positive and negative branches. The bias threshold at which this NDR occurs is independent of the magnetic flux (Fig. 3c) and scales parabolically with both in-plane magnetic fields and temperature (Fig. 3g-i). We attribute this behavior to the sudden loss of excess current as the superconducting leads are driven to their depairing limit. Thanks to the high transparency of the hBN-encapsulated graphene junction, Andreev reflections contribute an excess current to the $I-V$ characteristics even beyond the Josephson critical current. This contribution vanishes when the leads transition out of the superconducting state, triggering a sharp current drop as the voltage continues to rise, hence the observed NDR. In Fig. 3g, the asymptotic slopes of all $I\text{-}V$ curves extrapolate to the origin, signaling a loss of excess current already in the range between $1$ and $1.5 \; \mu$A, further corroborating our characterization of the Cu-doped-Al/Au bilayer leads.

\section*{Amplifier design and pair momentum detection}

The allowed range of $I_{\mathrm{ctrl}}$ is bound by the critical current of the superconducting leads. Expanding the accessible range of SGF to capture multiple oscillations in the SIP would conventionally require a higher critical current, which scales with the product of the lead cross-section and the Cooper-pair density $S n_s$. Meanwhile, a larger $Sn_s$ would suppress the kinetic inductance $L_k$, leaving the effective gauge field---proportional to $L_k I_{\mathrm{ctrl}}$---unchanged. This seemingly paradoxical constraint can be resolved by decoupling the phase-winding dimensions of the supercurrent and the junction. Notice the lead length governs the phase profile in Eqn. \eqref{lead_derivative}, whereas in Eqn. \eqref{ic_int}, the interference pattern is determined by integrating along the junction width. By engineering the superconducting leads into a meandering profile, we can accumulate an arbitrarily large phase difference from $\Delta L$ across the same lateral distance $\Delta x$ along the junction (Fig. 4a).

\begin{figure*}[htbp]
\centering
\includegraphics[width=0.7\textwidth]{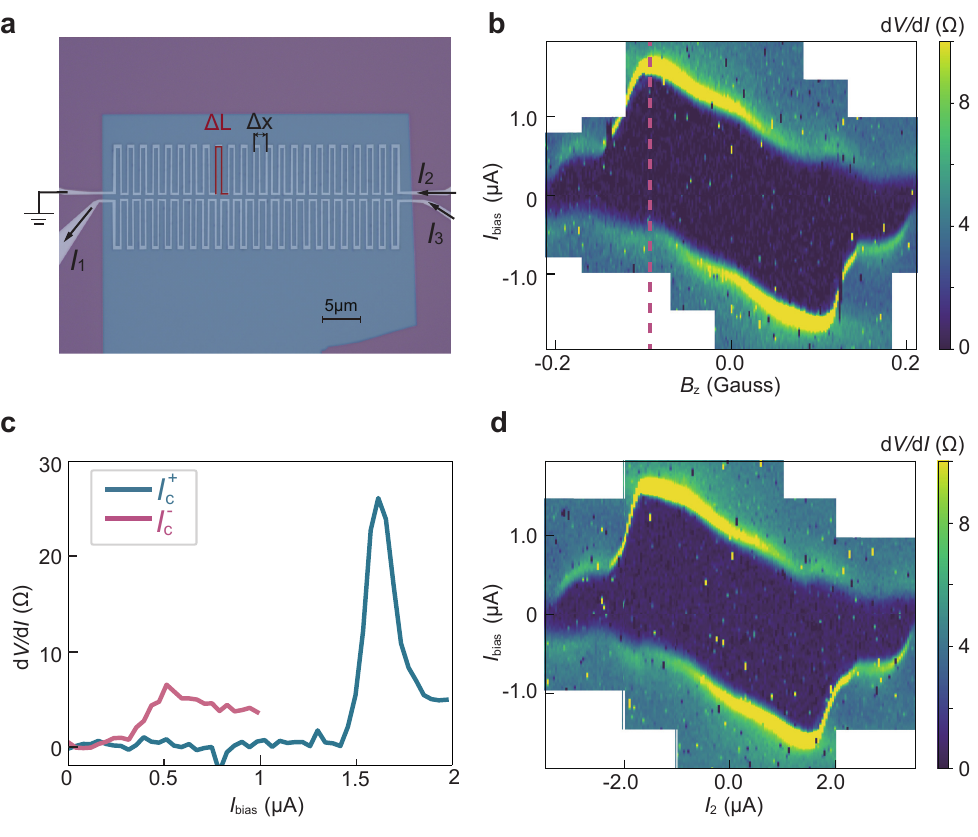} 
\caption{\textbf{Amplified SGF with meandering lead geometry.}
\textbf{a},  False-colour optical images of JJ2 with meandering leads. The electrode design (light gray) is overlaid on the optical micrograph to improve visibility. $\Delta L$ (red) and $\Delta x$ (black) mark each period along the lead length and junction length respectively. The meandering geometry yields a gauge amplification $\Delta L/\Delta x = 8.5$.
\textbf{b}, SIP under out-of-plane field $B_z$ (with $I_{3} = I_{2} = 0$ and $I_{\text{bias}} = -I_{1}$), showing a strongly skewed envelope and diode efficiencies up to 59\% (pink dashed line).
\textbf{c}, Linecut of differential resistance $dV/dI$ vs. bias current at the pink dashed line in \textbf{b}.
\textbf{d}, Magnetic-free SIP as a function of $I_{\mathrm{bias}}$ and $I_{2}$ at $B_z=0$, with $I_{3} = 0$ and $I_{\text{bias}} = -I_{1}$. The interference pattern and JDE magnitude closely resemble those in \textbf{b}, confirming that the SGF can fully replace an external magnetic field.}
\label{fig_4}
\end{figure*}

We implement this design with a gauge amplification of $8.5$, defined by the ratio $\Delta L / \Delta x$ (Fig. 4a). This geometric amplification enables the use of undoped Al within the Al/Au bilayer leads (Extended Data Fig. S5),  which provide substantially higher critical currents (exceeding $15\ \mu\mathrm{A}$). In the asymmetric configuration, the measured SIP in an external field (Fig. 4b) shows a dramatic skew, yielding diode efficiencies up to 59\% (Fig. 4c). At zero field, we show the full-electric-controlled SIP (Fig. 4d), which replicates the flux-induced pattern. The SIPs from the symmetric configuration are provided in Extended Data Fig. S6. These identical interference behaviors reveal on a profound level that ultimately, the role of magnetic flux is to induce phase frustration between the two sides of a Josephson junction—a frustration that can be agnostically achieved via either a magnetic flux or a SGF.

\begin{figure*}[htbp]
\centering
\includegraphics[width=0.9\textwidth]{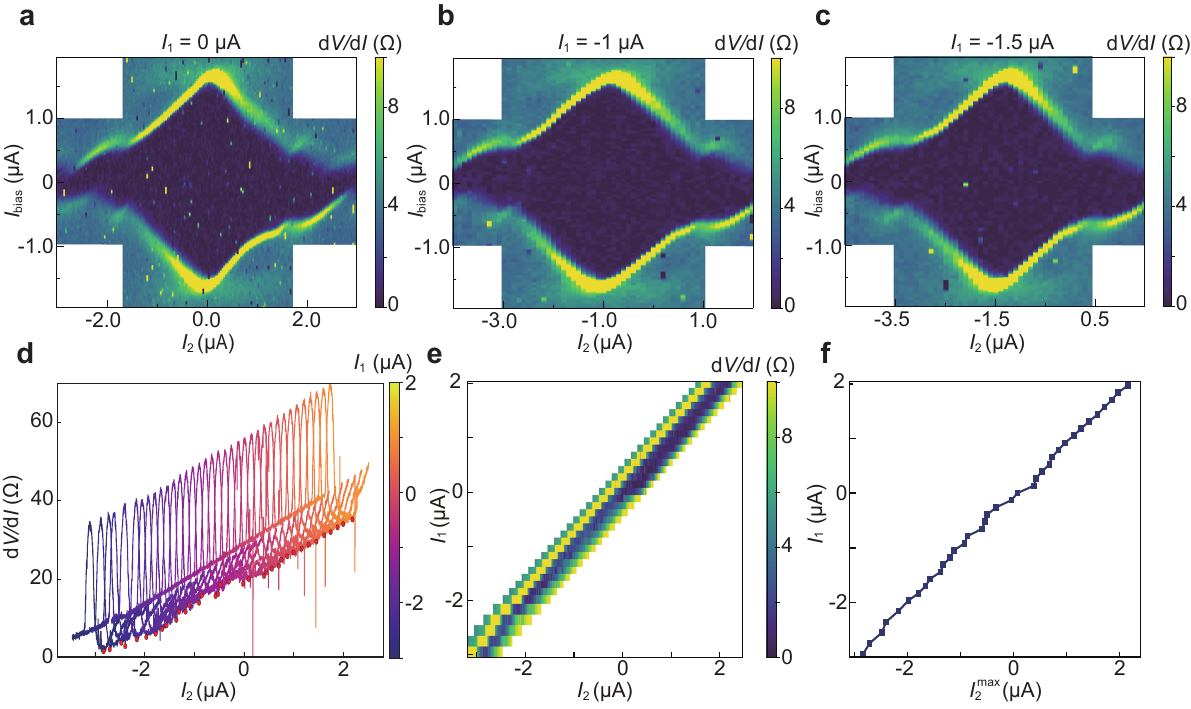} 
\caption{\textbf{Phase-sensitive generation and detection of finite-momentum superconductivity.}
\textbf{a–c}, SIPs showing differential resistance $dV/dI$ measured in JJ2 as a function of the control current $I_{2}$ and the bias current $I_{\text{bias}} = I_{3} - I_{1}$, measured at fixed $I_{1} = 0$ $\mu$A \textbf{a}, $-1$ $\mu$A \textbf{b}, and $-1.5$ $\mu$A \textbf{c}. The symmetric current configuration is used to suppress the universal JDE discussed in the main text. The maximum critical current occurs at $I_{2} = I_{1}$, the matching condition where the Cooper-pair momenta on the two sides are equal.
\textbf{d}, Waterfall plot of differential resistance $dV/dI$ vs. $I_{2}$ at $I_{\text{bias}} = 1.45\ \mu\text{A}$, recorded for different values of $I_{1}$. Red dots mark the extracted minima of the differential resistance, corresponding to the critical-current maxima. A $1\ \Omega$ relative offset is applied to each trace to enhance visibility.
\textbf{e}, The same data in \textbf{d} plotted as a colourmap.
\textbf{f}, Extracted positions of differential resistance minima vs. $I_{1}$ shows linear correlations.}
\label{fig_5}
\end{figure*}

Finally, we demonstrate the SGF offers local control flexibilities far beyond a uniform magnetic field. Operating with independent currents in both leads establishes two SGF knobs for supercurrent interference. When we apply a control current $I_i$ to lead $i$, it induces a Cooper-pair momentum equal to $L_k I_i$. Removing the magnetic term in Eqn. \eqref{phase_diff_x}, the total critical current becomes

\begin{equation} 
I_c =  \max_{\gamma_0} \int_{-W/2}^{W/2} J_c \sin(\gamma_0 + L_k (I_2 - I_1) x) \mathrm{d}x = I_{c0} \left| \mathrm{sinc} \left( \frac{L_k (I_2 - I_1) W }{2} \right) \right|,
\label{ic_finite-mmt}
\end{equation} 

which yields a Fraunhofer envelope featuring a sharp maximum centered right where the momenta on the two sides match. We obtain three interference patterns at $I_1 = 0, \; -1$, and $-1.5 \; \mu$A by sweeping $I_2$ (Fig. 5a-c). For these measurements, the bias current $I_{\mathrm{dc}}$ is sourced in the symmetric configuration to intentionally suppress the universal JDE described above. We can streamline this momentum-space mapping process by recording resistance linecuts at a high bias as a proxy for the full SIP evolution \autocite{Hart2016}. The results (Fig. 5d-e) exhibit a linear dependence of the resistance minima (critical-current maxima) on $I_1$. This demonstrates for the first time a quantum-transport method to perform spatial Fourier transform on a source of superconductivity with any finite momentum using, again, no magnetic field and only $\mu$A-level electric current.

\section*{Conclusion and outlook}

In conclusion, our results elucidate that the Josephson diode effect can be understood as a universal superconducting phase frustration---one that requires no intrinsic broken symmetry, spin-orbit coupling, or topological properties within the weak-link material itself. Enabled by the SGF, this JDE is an innate feature of the Josephson effect that has remained largely overlooked in prior literature due to the negligible kinetic inductance of traditional superconducting leads. While nonreciprocal transport has commonly been utilized as a definitive signature of broken symmetries in exotic states, our findings challenge this prevailing diagnostic paradigm for unconventional superconductivity, including finite-momentum pairing \autocite{Yuan2022, Davydova2022, Pal2022, Sivakumar2024}. At the time of finalizing this paper, we became aware of a preprint reporting similar phenomena in thin-film Nb junctions with $100 \; \mu$A-scale current and higher nonreciprocity \autocite{Hovhannisyan2026}. We would like to emphasize our work differs from this preprint and previous JDE discussions \autocite{Chen2024d, Guarcello2024} in that they still involve magnetic fields---whether externally applied or current-generated via self-fields---whereas our approach requires neither. By achieving all-electric control, we not only establish a universal magnetic-free JDE but also create a truly field-free paradigm to produce and manipulate supercurrent interference, which is well-suited for investigating real-space current distributions \autocite{Hart2014}, $4\pi$-periodic topological Josephson effects \autocite{Fu_2009}, and finite-momentum superconductivity \autocite{Hart2016, Pal2022}.

Moving forward, the conceptual groundwork and technology developed here open broad avenues for deterministic phase engineering in superconducting devices, equipping us with a cleaner, more versatile tool to break symmetries and pattern phases than traditional methods relying on magnetic fields or exchange interactions. For example, we can replace the source lead with superconductors of unknown pairing symmetries to study the real-space profiles of their order parameters. Note the meandering does not need to remain spatially uniform, nor are they restricted to linear geometries. This allows us the freedom to design arbitrary phase gradients in two- and three-dimensional quantum architectures. As an outlook, local vortex patterns on the surface states of topological insulators could host Majorana zero modes using currents alone \autocite{Fu2008, Melo2019, Lesser2021}. Most excitingly, integrating our SGF control with emerging quantum materials such as rhombohedral graphite \autocite{Patterson2025, Han2025}, twisted $\mathrm{MoTe}_2$ \autocite{Chen2026a}, and kagome metals \autocite{Zhong_2023, Le2024, Lou2026} offers a transformative, phase-sensitive methodology to probe, identify, and create exotic phases of matter.

\printbibliography

@Article{Zhong_2023,
  author    = {Yigui Zhong and Jinjin Liu and Xianxin Wu and Zurab Guguchia and J.-X. Yin and Akifumi Mine and Yongkai Li and Sahand Najafzadeh and Debarchan Das and Charles Mielke and Rustem Khasanov and Hubertus Luetkens and Takeshi Suzuki and Kecheng Liu and Xinloong Han and Takeshi Kondo and Jiangping Hu and Shik Shin and Zhiwei Wang and Xun Shi and Yugui Yao and Kozo Okazaki},
  journal   = {Nature},
  title     = {Nodeless electron pairing in {CsV$_3$Sb$_5$}-derived kagome superconductors},
  year      = {2023},
  month     = {apr},
  number    = {7961},
  pages     = {488--492},
  volume    = {617},
  doi       = {10.1038/s41586-023-05907-x},
  publisher = {Springer Science and Business Media {LLC}},
}

@Article{Fu_2009,
  author    = {Liang Fu and C. L. Kane},
  journal   = {Physical Review B},
  title     = {Josephson current and noise at a superconductor/quantum-spin-Hall-insulator/superconductor junction},
  year      = {2009},
  month     = {apr},
  number    = {16},
  pages     = {161408},
  volume    = {79},
  doi       = {10.1103/physrevb.79.161408},
  publisher = {American Physical Society ({APS})},
}

@Article{Pal2022,
  author    = {Pal, Banabir and Chakraborty, Anirban and Sivakumar, Pranava K. and Davydova, Margarita and Gopi, Ajesh K. and Pandeya, Avanindra K. and Krieger, Jonas A. and Zhang, Yang and Date, Mihir and Ju, Sailong and Yuan, Noah and Schröter, Niels B. M. and Fu, Liang and Parkin, Stuart S. P.},
  journal   = {Nature Physics},
  title     = {Josephson diode effect from Cooper pair momentum in a topological semimetal},
  year      = {2022},
  issn      = {1745-2481},
  month     = {aug},
  number    = {10},
  pages     = {1228--1233},
  volume    = {18},
  doi       = {10.1038/s41567-022-01699-5},
  publisher = {Springer Science and Business Media LLC},
}

@Article{Ren2019,
  author    = {Hechen Ren and Falko Pientka and Sean Hart and Andrew T. Pierce and Michael Kosowsky and Lukas Lunczer and Raimund Schlereth and Benedikt Scharf and Ewelina M. Hankiewicz and Laurens W. Molenkamp and Bertrand I. Halperin and Amir Yacoby},
  journal   = {Nature},
  title     = {Topological superconductivity in a phase-controlled Josephson junction},
  year      = {2019},
  issn      = {1476-4687},
  month     = {apr},
  number    = {7754},
  pages     = {93--98},
  volume    = {569},
  doi       = {10.1038/s41586-019-1148-9},
  publisher = {Springer Science and Business Media {LLC}},
}

@Article{Hart2016,
  author    = {Hart, Sean and Ren, Hechen and Kosowsky, Michael and Ben-Shach, Gilad and Leubner, Philipp and Brüne, Christoph and Buhmann, Hartmut and Molenkamp, Laurens W. and Halperin, Bertrand I. and Yacoby, Amir},
  journal   = {Nature Physics},
  title     = {Controlled finite momentum pairing and spatially varying order parameter in proximitized {HgTe} quantum wells},
  year      = {2016},
  issn      = {1745-2481},
  month     = {sep},
  number    = {1},
  pages     = {87--93},
  volume    = {13},
  doi       = {10.1038/nphys3877},
  publisher = {Springer Science and Business Media LLC},
}

@Article{Hart2014,
  author    = {Hart, Sean and Ren, Hechen and Wagner, Timo and Leubner, Philipp and Mühlbauer, Mathias and Brüne, Christoph and Buhmann, Hartmut and Molenkamp, Laurens W. and Yacoby, Amir},
  journal   = {Nature Physics},
  title     = {Induced superconductivity in the quantum spin Hall edge},
  year      = {2014},
  issn      = {1745-2481},
  month     = {aug},
  number    = {9},
  pages     = {638--643},
  volume    = {10},
  doi       = {10.1038/nphys3036},
  publisher = {Springer Science and Business Media LLC},
}

@Article{Davydova2022,
  author    = {Davydova, Margarita and Prembabu, Saranesh and Fu, Liang},
  journal   = {Science Advances},
  title     = {Universal Josephson diode effect},
  year      = {2022},
  issn      = {2375-2548},
  month     = {jun},
  number    = {23},
  volume    = {8},
  pages     = {eabo0309},
  doi       = {10.1126/sciadv.abo0309},
  publisher = {American Association for the Advancement of Science (AAAS)},
}

@Article{Gupta2023a,
  author    = {Gupta, Mohit and Graziano, Gino V. and Pendharkar, Mihir and Dong, Jason T. and Dempsey, Connor P. and Palmstrøm, Chris and Pribiag, Vlad S.},
  journal   = {Nature Communications},
  title     = {Gate-tunable superconducting diode effect in a three-terminal Josephson device},
  year      = {2023},
  issn      = {2041-1723},
  month     = {may},
  number    = {1},
  volume    = {14},
  pages     = {3078},
  doi       = {10.1038/s41467-023-38856-0},
  publisher = {Springer Science and Business Media LLC},
}

@Article{Kraft2018,
  author    = {Kraft, Rainer and Mohrmann, Jens and Du, Renjun and Selvasundaram, Pranauv Balaji and Irfan, Muhammad and Kanilmaz, Umut Nefta and Wu, Fan and Beckmann, Detlef and von Löhneysen, Hilbert and Krupke, Ralph and Akhmerov, Anton and Gornyi, Igor and Danneau, Romain},
  journal   = {Nature Communications},
  title     = {Tailoring supercurrent confinement in graphene bilayer weak links},
  year      = {2018},
  issn      = {2041-1723},
  month     = {apr},
  number    = {1},
  volume    = {9},
  doi       = {10.1038/s41467-018-04153-4},
  pages     = {1722},
  publisher = {Springer Science and Business Media LLC},
}

@Article{Rout2024,
  author    = {Rout, Prasanna and Papadopoulos, Nikos and Peñaranda, Fernando and Watanabe, Kenji and Taniguchi, Takashi and Prada, Elsa and San-Jose, Pablo and Goswami, Srijit},
  journal   = {Nature Communications},
  title     = {Supercurrent mediated by helical edge modes in bilayer graphene},
  year      = {2024},
  issn      = {2041-1723},
  month     = {jan},
  number    = {1},
  volume    = {15},
  pages     = {856},
  doi       = {10.1038/s41467-024-44952-6},
  publisher = {Springer Science and Business Media LLC},
}

@Article{Patterson2025,
  author    = {Patterson, Caitlin L. and Sheekey, Owen I. and Arp, Trevor B. and Holleis, Ludwig F. W. and Koh, Jin Ming and Choi, Youngjoon and Xie, Tian and Xu, Siyuan and Guo, Yi and Stoyanov, Hari and Redekop, Evgeny and Zhang, Canxun and Babikyan, Grigory and Gong, David and Zhou, Haoxin and Cheng, Xiang and Taniguchi, Takashi and Watanabe, Kenji and Huber, Martin E. and Jin, Chenhao and Lantagne-Hurtubise, {\'E}tienne and Alicea, Jason and Young, Andrea F.},
  journal   = {Nature},
  title     = {Superconductivity and spin canting in spin–orbit-coupled trilayer graphene},
  year      = {2025},
  issn      = {1476-4687},
  month     = {may},
  number    = {8063},
  pages     = {632--638},
  volume    = {641},
  doi       = {10.1038/s41586-025-08863-w},
  publisher = {Springer Science and Business Media LLC},
}

@Article{Sivakumar2024,
  author    = {Sivakumar, Pranava K. and Ahari, Mostafa T. and Kim, Jae-Keun and Wu, Yufeng and Dixit, Anvesh and de Coster, George J. and Pandeya, Avanindra K. and Gilbert, Matthew J. and Parkin, Stuart S. P.},
  journal   = {Communications Physics},
  title     = {Long-range phase coherence and tunable second order $\varphi_0$-Josephson effect in a Dirac semimetal {1T-PtTe$_2$}},
  year      = {2024},
  issn      = {2399-3650},
  month     = {oct},
  number    = {1},
  volume    = {7},
  doi       = {10.1038/s42005-024-01825-0},
  pages     = {354},
  publisher = {Springer Science and Business Media LLC},
}

@Article{Ma2025a,
  author    = {Ma, Jiaxiang and Wang, Huiyu and Zhuo, Weizhuang and Lei, Bin and Wang, Shuai and Wang, Wenxiang and Chen, Xin-Yu and Wang, Zhen-Yu and Ge, Binghui and Wang, Zhen and Tao, Jing and Jiang, Kun and Xiang, Ziji and Chen, Xian-Hui},
  journal   = {Communications Physics},
  title     = {Field-free Josephson diode effect in {NbSe$_2$} van der Waals junction},
  year      = {2025},
  issn      = {2399-3650},
  month     = {mar},
  number    = {1},
  volume    = {8},
  pages     = {125},
  doi       = {10.1038/s42005-025-02054-9},
  publisher = {Springer Science and Business Media LLC},
}

@Article{Wu2022a,
  author    = {Wu, Heng and Wang, Yaojia and Xu, Yuanfeng and Sivakumar, Pranava K. and Pasco, Chris and Filippozzi, Ulderico and Parkin, Stuart S. P. and Zeng, Yu-Jia and McQueen, Tyrel and Ali, Mazhar N.},
  journal   = {Nature},
  title     = {The field-free Josephson diode in a van der Waals heterostructure},
  year      = {2022},
  issn      = {1476-4687},
  month     = apr,
  number    = {7907},
  pages     = {653--656},
  volume    = {604},
  doi       = {10.1038/s41586-022-04504-8},
  publisher = {Springer Science and Business Media LLC},
}

@Article{Jeon2022,
  author    = {Jeon, Kun-Rok and Kim, Jae-Keun and Yoon, Jiho and Jeon, Jae-Chun and Han, Hyeon and Cottet, Audrey and Kontos, Takis and Parkin, Stuart S. P.},
  journal   = {Nature Materials},
  title     = {Zero-field polarity-reversible Josephson supercurrent diodes enabled by a proximity-magnetized {Pt} barrier},
  year      = {2022},
  issn      = {1476-4660},
  month     = jul,
  number    = {9},
  pages     = {1008--1013},
  volume    = {21},
  doi       = {10.1038/s41563-022-01300-7},
  publisher = {Springer Science and Business Media LLC},
}

@Article{Han2025,
  author    = {Han, Tonghang and Lu, Zhengguang and Hadjri, Zach and Shi, Lihan and Wu, Zhenghan and Xu, Wei and Yao, Yuxuan and Cotten, Armel A. and Sharifi Sedeh, Omid and Weldeyesus, Henok and Yang, Jixiang and Seo, Junseok and Ye, Shenyong and Zhou, Muyang and Liu, Haoyang and Shi, Gang and Hua, Zhenqi and Watanabe, Kenji and Taniguchi, Takashi and Xiong, Peng and Zumbühl, Dominik M. and Fu, Liang and Ju, Long},
  journal   = {Nature},
  title     = {Signatures of chiral superconductivity in rhombohedral graphene},
  year      = {2025},
  issn      = {1476-4687},
  month     = may,
  number    = {8072},
  pages     = {654--661},
  volume    = {643},
  doi       = {10.1038/s41586-025-09169-7},
  publisher = {Springer Science and Business Media LLC},
}

@Article{Golod2022,
  author    = {Golod, Taras and Krasnov, Vladimir M.},
  journal   = {Nature Communications},
  title     = {Demonstration of a superconducting diode-with-memory, operational at zero magnetic field with switchable nonreciprocity},
  year      = {2022},
  issn      = {2041-1723},
  month     = jun,
  number    = {1},
  volume    = {13},
  pages     = {3658},
  doi       = {10.1038/s41467-022-31256-w},
  publisher = {Springer Science and Business Media LLC},
}

@Article{Hou2023,
  author    = {Hou, Yasen and Nichele, Fabrizio and Chi, Hang and Lodesani, Alessandro and Wu, Yingying and Ritter, Markus F. and Haxell, Daniel Z. and Davydova, Margarita and Ilić, Stefan and Glezakou-Elbert, Ourania and Varambally, Amith and Bergeret, F. Sebastian and Kamra, Akashdeep and Fu, Liang and Lee, Patrick A. and Moodera, Jagadeesh S.},
  journal   = {Physical Review Letters},
  title     = {Ubiquitous Superconducting Diode Effect in Superconductor Thin Films},
  year      = {2023},
  issn      = {1079-7114},
  month     = jul,
  number    = {2},
  pages     = {027001},
  volume    = {131},
  doi       = {10.1103/physrevlett.131.027001},
  publisher = {American Physical Society (APS)},
}

@Article{Le2024,
  author    = {Le, Tian and Pan, Zhiming and Xu, Zhuokai and Liu, Jinjin and Wang, Jialu and Lou, Zhefeng and Yang, Xiaohui and Wang, Zhiwei and Yao, Yugui and Wu, Congjun and Lin, Xiao},
  journal   = {Nature},
  title     = {Superconducting diode effect and interference patterns in kagome {CsV$_3$Sb$_5$}},
  year      = {2024},
  issn      = {1476-4687},
  month     = may,
  number    = {8015},
  pages     = {64--69},
  volume    = {630},
  doi       = {10.1038/s41586-024-07431-y},
  publisher = {Springer Science and Business Media LLC},
  ranking   = {rank5},
}

@Article{Chiles2023,
  author    = {Chiles, John and Arnault, Ethan G. and Chen, Chun-Chia and Larson, Trevyn F. Q. and Zhao, Lingfei and Watanabe, Kenji and Taniguchi, Takashi and Amet, François and Finkelstein, Gleb},
  journal   = {Nano Letters},
  title     = {Nonreciprocal Supercurrents in a Field-Free Graphene Josephson Triode},
  year      = {2023},
  issn      = {1530-6992},
  month     = may,
  number    = {11},
  pages     = {5257--5263},
  volume    = {23},
  doi       = {10.1021/acs.nanolett.3c01276},
  publisher = {American Chemical Society (ACS)},
}

@Article{Lesser2024,
  author    = {Lesser, Omri and Stern, Ady and Oreg, Yuval},
  journal   = {Physical Review B},
  title     = {Josephson junction arrays as a platform for topological phases of matter},
  year      = {2024},
  issn      = {2469-9969},
  month     = apr,
  number    = {14},
  pages     = {144519},
  volume    = {109},
  doi       = {10.1103/physrevb.109.144519},
  publisher = {American Physical Society (APS)},
}

@Article{Loo2026,
  author    = {van Loo, Nick and Zatelli, Francesco and Steffensen, Gorm O. and Roovers, Bart and Wang, Guanzhong and Van Caekenberghe, Thomas and Bordin, Alberto and van Driel, David and Zhang, Yining and Huisman, Wietze D. and Badawy, Ghada and Bakkers, Erik P. A. M. and Mazur, Grzegorz P. and Aguado, Ramón and Kouwenhoven, Leo P.},
  journal   = {Nature},
  title     = {Single-shot parity readout of a minimal Kitaev chain},
  year      = {2026},
  issn      = {1476-4687},
  month     = feb,
  number    = {8101},
  pages     = {334--339},
  volume    = {650},
  doi       = {10.1038/s41586-025-09927-7},
  publisher = {Springer Science and Business Media LLC},
}

@Article{Hu2020,
  author    = {Hu, Jie and Salatino, Maria and Traini, Alessandro and Chaumont, Christine and Boussaha, Faouzi and Goupil, Christophe and Piat, Michel},
  journal   = {Journal of Low Temperature Physics},
  title     = {Proximity-Coupled {Al/Au} Bilayer Kinetic Inductance Detectors},
  year      = {2020},
  issn      = {1573-7357},
  month     = jan,
  number    = {1–2},
  pages     = {355--361},
  volume    = {199},
  doi       = {10.1007/s10909-019-02313-4},
  publisher = {Springer Science and Business Media LLC},
}

@Article{Baumgartner2021,
  author    = {Baumgartner, Christian and Fuchs, Lorenz and Costa, Andreas and Reinhardt, Simon and Gronin, Sergei and Gardner, Geoffrey C. and Lindemann, Tyler and Manfra, Michael J. and Faria Junior, Paulo E. and Kochan, Denis and Fabian, Jaroslav and Paradiso, Nicola and Strunk, Christoph},
  journal   = {Nature Nanotechnology},
  title     = {Supercurrent rectification and magnetochiral effects in symmetric Josephson junctions},
  year      = {2021},
  issn      = {1748-3395},
  month     = nov,
  number    = {1},
  pages     = {39--44},
  volume    = {17},
  doi       = {10.1038/s41565-021-01009-9},
  publisher = {Springer Science and Business Media LLC},
}

@Article{Wakatsuki2017,
  author    = {Wakatsuki, Ryohei and Saito, Yu and Hoshino, Shintaro and Itahashi, Yuki M. and Ideue, Toshiya and Ezawa, Motohiko and Iwasa, Yoshihiro and Nagaosa, Naoto},
  journal   = {Science Advances},
  title     = {Nonreciprocal charge transport in noncentrosymmetric superconductors},
  year      = {2017},
  issn      = {2375-2548},
  month     = apr,
  number    = {4},
  volume    = {3},
  doi       = {10.1126/sciadv.1602390},
  pages     = {e1602390},
  publisher = {American Association for the Advancement of Science (AAAS)},
}

@Article{Ando2020,
  author    = {Ando, Fuyuki and Miyasaka, Yuta and Li, Tian and Ishizuka, Jun and Arakawa, Tomonori and Shiota, Yoichi and Moriyama, Takahiro and Yanase, Youichi and Ono, Teruo},
  journal   = {Nature},
  title     = {Observation of superconducting diode effect},
  year      = {2020},
  issn      = {1476-4687},
  month     = aug,
  number    = {7821},
  pages     = {373--376},
  volume    = {584},
  doi       = {10.1038/s41586-020-2590-4},
  publisher = {Springer Science and Business Media LLC},
}

@Article{Wang2023c,
  author    = {Wang, Gensheng and Barry, Peter S. and Cecil, Thomas and Chang, Clarence L. and Li, Juliang and Lisovenko, Marharyta and Novosad, Valentyn and Pan, Zhaodi and Yefremenko, Volodymyr G. and Zhang, Jianjie},
  journal   = {IEEE Transactions on Applied Superconductivity},
  title     = {Electromagnetic Properties of Aluminum-Based Bilayers for Kinetic Inductance Detectors},
  year      = {2023},
  issn      = {2378-7074},
  month     = Aug,
  number    = {5},
  pages     = {1--6},
  volume    = {33},
  doi       = {10.1109/tasc.2023.3245059},
  publisher = {Institute of Electrical and Electronics Engineers (IEEE)},
}

@Article{DiezMerida2023,
  author    = {Díez-Mérida, J. and Díez-Carlón, A. and Yang, S. Y. and Xie, Y.-M. and Gao, X.-J. and Senior, J. and Watanabe, K. and Taniguchi, T. and Lu, X. and Higginbotham, A. P. and Law, K. T. and Efetov, Dmitri K.},
  journal   = {Nature Communications},
  title     = {Symmetry-broken Josephson junctions and superconducting diodes in magic-angle twisted bilayer graphene},
  year      = {2023},
  issn      = {2041-1723},
  month     = apr,
  number    = {1},
  volume    = {14},
  doi       = {10.1038/s41467-023-38005-7},
  pages     = {2396},
  publisher = {Springer Science and Business Media LLC},
}

@Article{Martinis1985,
  author    = {Martinis, John M. and Devoret, Michel H. and Clarke, John},
  journal   = {Physical Review Letters},
  title     = {Energy-Level Quantization in the Zero-Voltage State of a Current-Biased Josephson Junction},
  year      = {1985},
  issn      = {0031-9007},
  month     = Oct,
  number    = {15},
  pages     = {1543--1546},
  volume    = {55},
  doi       = {10.1103/physrevlett.55.1543},
  publisher = {American Physical Society (APS)},
}

@Article{Devoret1985,
  author    = {Devoret, Michel H. and Martinis, John M. and Clarke, John},
  journal   = {Physical Review Letters},
  title     = {Measurements of Macroscopic Quantum Tunneling out of the Zero-Voltage State of a Current-Biased Josephson Junction},
  year      = {1985},
  issn      = {0031-9007},
  month     = oct,
  number    = {18},
  pages     = {1908--1911},
  volume    = {55},
  doi       = {10.1103/physrevlett.55.1908},
  publisher = {American Physical Society (APS)},
}

@Article{Dobrovolskiy2026,
  author    = {Dobrovolskiy, Oleksandr and Suderow, Hermann and Tafuri, Francesco and Black-Schaffer, Annica M and Lado, Jose L and Sudbø, Asle and Stornaioulo, Daniela and Li, Chuan and Böhmer, Anna E and Tran, Lan Maria and Zaleski, Andrzej J and Crisan, Adrian and Polichetti, Massimiliano and Galluzzi, Armando and Gencer, Ali and Aichner, Bernd and Barišić, Neven and Lang, Wolfgang and Samuely, Tomas and Gmitra, Martin and Cren, Tristan and Calandra, Mateo and Samuely, Peter and Custers, Jeroen and Córdoba, Rosa and Fomin, Vladimir M and Poccia, Nicola and Szabó, Pavol and Porrati, Fabrizio and Kakazei, Gleb and Aarts, Jan and Robinson, Jason and Villegas, Javier E and Althammer, Matthias and Huebl, Hans and Kamra, Akashdeep and Weiler, Mathias and Dil, J Hugo and Evtushinsky, Daniil and Kalisky, Beena and Anahory, Yonathan and Bending, Simon and Liljeroth, Peter and Hassanien, Abdou and Guillamón, Isabel and Herrera, Edwin and Silhanek, Alejandro V and Van de Vondel, Joris and Palau, Anna and Charaev, Ilya and Sidorova, Maria and Lombardi, Floriana and Bauch, Thilo and Feuillet-Palma, Cheryl and Stolyarov, Vasily and Roditchev, Dimitri and Krasnov, Vladimir M and Hampel, Benedikt and Martínez-Pérez, María José and Sesé, Javier and Koelle, Dieter and Poletto, Stefano and Bruno, Alessandro and Massarotti, Davide},
  journal   = {Superconductor Science and Technology},
  title     = {Roadmap on nanoscale superconductivity for quantum technologies},
  year      = {2026},
  issn      = {1361-6668},
  month     = Feb,
  number    = {2},
  pages     = {023502},
  volume    = {39},
  doi       = {10.1088/1361-6668/ae3030},
  publisher = {IOP Publishing},
}

@Article{Kim2024,
  author    = {Kim, Jae-Keun and Jeon, Kun-Rok and Sivakumar, Pranava K. and Jeon, Jaechun and Koerner, Chris and Woltersdorf, Georg and Parkin, Stuart S. P.},
  journal   = {Nature Communications},
  title     = {Intrinsic supercurrent non-reciprocity coupled to the crystal structure of a van der Waals Josephson barrier},
  year      = {2024},
  issn      = {2041-1723},
  month     = Feb,
  number    = {1},
  volume    = {15},
  doi       = {10.1038/s41467-024-45298-9},
  pages     = {1120},
  publisher = {Springer Science and Business Media LLC},
}

@Article{Chen2026a,
  author    = {Chen, Yinqi and Xu, Cheng and Zhang, Yang and Schrade, Constantin},
  journal   = {Nature Communications},
  title     = {Finite-momentum superconductivity from chiral bands in twisted {MoTe$_2$}},
  year      = {2026},
  issn      = {2041-1723},
  month     = {Jan},
  pages = {1077},
  number    = {1},
  volume    = {17},
  doi       = {10.1038/s41467-025-67836-9},
  publisher = {Springer Science and Business Media LLC},
}

@Article{Yuan2022,
  author    = {Yuan, Noah F. Q. and Fu, Liang},
  journal   = {Proceedings of the National Academy of Sciences},
  title     = {Supercurrent diode effect and finite-momentum superconductors},
  year      = {2022},
  issn      = {1091-6490},
  month     = {apr},
  number    = {15},
  volume    = {119},
  doi       = {10.1073/pnas.2119548119},
  pages     = {e2119548119},
  publisher = {Proceedings of the National Academy of Sciences},
}

@Article{Lesser2021,
  author    = {Lesser, Omri and Saydjari, Andrew and Wesson, Marie and Yacoby, Amir and Oreg, Yuval},
  journal   = {Proceedings of the National Academy of Sciences},
  title     = {Phase-induced topological superconductivity in a planar heterostructure},
  year      = {2021},
  issn      = {1091-6490},
  month     = {June},
  number    = {27},
  volume    = {118},
  doi       = {10.1073/pnas.2107377118},
  pages     = {e2107377118},
  publisher = {Proceedings of the National Academy of Sciences},
}

@Article{Ying2020,
  author    = {Ying, Jianghua and He, Jiangbo and Yang, Guang and Liu, Mingli and Lyu, Zhaozheng and Zhang, Xiang and Liu, Huaiyuan and Zhao, Kui and Jiang, Ruiyang and Ji, Zhongqing and Fan, Jie and Yang, Changli and Jing, Xiunian and Liu, Guangtong and Cao, Xuewei and Wang, Xuefeng and Lu, Li and Qu, Fanming},
  journal   = {Nano Letters},
  title     = {Magnitude and Spatial Distribution Control of the Supercurrent in {Bi$_2$O$_2$Se}-Based Josephson Junction},
  year      = {2020},
  issn      = {1530-6992},
  month     = {Mar},
  number    = {4},
  pages     = {2569--2575},
  volume    = {20},
  doi       = {10.1021/acs.nanolett.0c00025},
  publisher = {American Chemical Society (ACS)},
}

@Article{Lou2026,
  author    = {Lou, Han-Xin and Chen, Jing-Jing and Ye, Xing-Guo and Tan, Zhen-Bing and Wang, An-Qi and Yin, Qing and Liao, Xin and Fang, Jing-Zhi and Liu, Xing-Yu and He, Yi-Lin and Zhang, Zhen-Tao and Li, Chuan and Wei, Zhong-Ming and Ma, Xiu-Mei and Yu, Da-Peng and Liao, Zhi-Min},
  journal   = {Nature Nanotechnology},
  title     = {Quantized radio-frequency rectification in a kagome superconductor Josephson diode},
  year      = {2026},
  issn      = {1748-3395},
  month     = Jan,
  number    = {4},
  pages     = {554--560},
  volume    = {21},
  doi       = {10.1038/s41565-025-02120-x},
  publisher = {Springer Science and Business Media LLC},
}

@Article{Trahms2023,
  author    = {Martina Trahms and Larissa Melischek and Jacob F. Steiner and Bharti Mahendru and Idan Tamir and Nils Bogdanoff and Olof Peters and Gaël Reecht and Clemens B. Winkelmann and Felix von Oppen and Katharina J. Franke},
  journal   = {Nature},
  title     = {Diode effect in Josephson junctions with a single magnetic atom},
  year      = {2023},
  issn      = {1476-4687},
  month     = {mar},
  number    = {7953},
  pages     = {628--633},
  volume    = {615},
  doi       = {10.1038/s41586-023-05743-z},
  publisher = {Springer Science and Business Media {LLC}},
}

@Article{Zhu2025,
  author    = {Zhu, Shang and Ma, Yiwen and He, Jiangbo and Yang, Xiaozhou and Jia, Zhongmou and Wei, Min and Jiao, Yiping and He, Jiezhong and Zhuo, Enna and Cao, Xuewei and Tong, Bingbing and Dou, Ziwei and Li, Peiling and Shen, Jie and Song, Xiaohui and Lyu, Zhaozheng and Liu, Guangtong and Pan, Dong and Zhao, Jianhua and Lu, Bo and Lu, Li and Qu, Fanming},
  journal   = {Communications Physics},
  title     = {Josephson diode effect in nanowire-based Andreev molecules},
  year      = {2025},
  issn      = {2399-3650},
  month     = aug,
  number    = {1},
  volume    = {8},
  doi       = {10.1038/s42005-025-02237-4},
  pages     = {330},
  publisher = {Springer Science and Business Media LLC},
}

@Article{Matsuo2023a,
  author    = {Matsuo, Sadashige and Imoto, Takaya and Yokoyama, Tomohiro and Sato, Yosuke and Lindemann, Tyler and Gronin, Sergei and Gardner, Geoffrey C. and Manfra, Michael J. and Tarucha, Seigo},
  journal   = {Nature Physics},
  title     = {Josephson diode effect derived from short-range coherent coupling},
  year      = {2023},
  issn      = {1745-2481},
  month     = July,
  number    = {11},
  pages     = {1636--1641},
  volume    = {19},
  doi       = {10.1038/s41567-023-02144-x},
  publisher = {Springer Science and Business Media LLC},
}

@Article{Chen2024d,
  author    = {Chen, Shaowen and Park, Seunghyun and Vool, Uri and Maksimovic, Nikola and Broadway, David A. and Flaks, Mykhailo and Zhou, Tony X. and Maletinsky, Patrick and Stern, Ady and Halperin, Bertrand I. and Yacoby, Amir},
  title     = {Current induced hidden states in Josephson junctions},
  journal   = {\href{https://doi.org/10.1038/s41467-024-52271-z}{Nature Communications}},
  year      = {2024},
  volume    = {15},
  number    = {1},
  issn      = {2041-1723},
  pages     = {8059},
  month     = Sept, 
  doi       = {10.1038/s41467-024-52271-z},
}

@Book{Tinkham2004,
  author    = {Tinkham, M.},
  publisher = {Dover Publications},
  title     = {Introduction to Superconductivity},
  year      = {2004},
  isbn      = {9780486134727},
  series    = {Dover Books on Physics Series},
  url       = {https://books.google.com/books?id=VpUk3NfwDIkC},
}

@Article{Guarcello2024,
  author    = {Guarcello, C. and Pagano, S. and Filatrella, G.},
  journal   = {Applied Physics Letters},
  title     = {Efficiency of diode effect in asymmetric inline long Josephson junctions},
  year      = {2024},
  issn      = {1077-3118},
  month     = apr,
  number    = {16},
  volume    = {124},
  doi       = {10.1063/5.0211230},
  pages     = {162601},
  publisher = {AIP Publishing},
}

@Article{Melo2019,
  author    = {Melo, André and Rubbert, Sebastian and Akhmerov, Anton R.},
  journal   = {SciPost Physics},
  title     = {Supercurrent-induced Majorana bound states in a planar geometry},
  year      = {2019},
  issn      = {2542-4653},
  month     = Sept,
  number    = {3},
  volume    = {7},
  doi       = {10.21468/scipostphys.7.3.039},
  pages = {039},
  publisher = {Stichting SciPost},
}

@Article{Nguyen2019,
  author    = {Nguyen, Long B. and Lin, Yen-Hsiang and Somoroff, Aaron and Mencia, Raymond and Grabon, Nicholas and Manucharyan, Vladimir E.},
  journal   = {Physical Review X},
  title     = {High-Coherence Fluxonium Qubit},
  year      = {2019},
  issn      = {2160-3308},
  month     = Nov,
  number    = {4},
  pages     = {041041},
  volume    = {9},
  doi       = {10.1103/physrevx.9.041041},
  publisher = {American Physical Society (APS)},
}

@Misc{Fu2026,
  author    = {Fu, J. B. and Wang, Da-Wei and Ren, B. and Yang, Z. H. and Hu, S. and Huang, G. Y. and Cao, S. H. and Liu, D. D. and Zhang, X. F. and Fu, X. and Xue, S. C. and Che, Y. G. and Liu, Yu-xi and Deng, M. T. and Wu, J. J.},
  title     = {Flux-noise-resilient transmon qubit via a doubly-connected gradiometric design},
  year      = {2026},
  copyright = {Creative Commons Attribution 4.0 International},
  eprint        = {2601.02137},
  archivePrefix = {arXiv},
  publisher = {arXiv},
}

@Article{Rower2023,
  author    = {Rower, David A. and Ateshian, Lamia and Li, Lauren H. and Hays, Max and Bluvstein, Dolev and Ding, Leon and Kannan, Bharath and Almanakly, Aziza and Braumüller, Jochen and Kim, David K. and Melville, Alexander and Niedzielski, Bethany M. and Schwartz, Mollie E. and Yoder, Jonilyn L. and Orlando, Terry P. and Wang, Joel I-Jan and Gustavsson, Simon and Grover, Jeffrey A. and Serniak, Kyle and Comin, Riccardo and Oliver, William D.},
  journal   = {Physical Review Letters},
  title     = {Evolution of 1/f Flux Noise in Superconducting Qubits with Weak Magnetic Fields},
  year      = {2023},
  issn      = {1079-7114},
  month     = May,
  number    = {22},
  pages     = {220602},
  volume    = {130},
  doi       = {10.1103/physrevlett.130.220602},
  publisher = {American Physical Society (APS)},
}

@Article{Valenti2019,
  author    = {Valenti, Francesco and Henriques, Fabio and Catelani, Gianluigi and Maleeva, Nataliya and Grünhaupt, Lukas and von Lüpke, Uwe and Skacel, Sebastian T. and Winkel, Patrick and Bilmes, Alexander and Ustinov, Alexey V. and Goupy, Johannes and Calvo, Martino and Benoît, Alain and Levy-Bertrand, Florence and Monfardini, Alessandro and Pop, Ioan M.},
  journal   = {Physical Review Applied},
  title     = {Interplay Between Kinetic Inductance, Nonlinearity, and Quasiparticle Dynamics in Granular Aluminum Microwave Kinetic Inductance Detectors},
  year      = {2019},
  issn      = {2331-7019},
  month     = {May},
  number    = {5},
  pages     = {054087},
  volume    = {11},
  doi       = {10.1103/physrevapplied.11.054087},
  publisher = {American Physical Society (APS)},
}

@Article{Fu2008,
  author    = {Liang Fu and C. L. Kane},
  journal   = {Physical Review Letters},
  title     = {Superconducting Proximity Effect and Majorana Fermions at the Surface of a Topological Insulator},
  year      = {2008},
  issn      = {1079-7114},
  month     = {mar},
  number    = {9},
  pages     = {096407},
  volume    = {100},
  doi       = {10.1103/physrevlett.100.096407},
  publisher = {American Physical Society ({APS})},
}

@Misc{Hovhannisyan2026,
  author    = {Hovhannisyan, Razmik A. and Golod, Taras and Lotfian, Amirreza and Krasnov, Vladimir M.},
  title     = {In-situ tunable superconducting diode: towards field-free operation with infinite nonreciprocity},
  year      = {2026},
  copyright = {Creative Commons Zero v1.0 Universal},
  eprint        = {2605.13254},
  archivePrefix = {arXiv},
  publisher = {arXiv},
}

@article{Liu2020,
author = {Hong, Chengyun and Tao, Ye and Nie, Anmin and Zhang, Minhao and Wang, Nan and Li, Ruiping and Huang, Junquan and Huang, Yongqing and Ren, Xiaomin and Cheng, Yingchun and Liu, Xiaolong},
title = {Inclined Ultrathin {Bi$_2$O$_2$Se} Films: A Building Block for Functional van der Waals Heterostructures},
journal = {ACS Nano},
volume = {14},
number = {12},
pages = {16803-16812},
year = {2020},
doi = {10.1021/acsnano.0c05300},
}

\clearpage

\section*{Methods}

\textbf{Sample fabrication and transport measurement}

$\mathrm{Bi}_2\mathrm{O}_2\mathrm{Se}$ was synthesized via chemical vapor deposition (CVD) in a dual-zone tube furnace. In the upstream zone, the $\mathrm{Bi}_2\mathrm{Se}_3$ precursor (3A, 99.999\%) was placed at the hot center. In the downstream zone, the $\mathrm{Bi}_2\mathrm{O}_3$ precursor (Alfa Aesar, 99.995\%) was positioned $7$--$10$~cm away from the hot center. Fluorophlogopite mica substrates were placed directly above the $\mathrm{Bi}_2\mathrm{O}_3$ source. During the growth, the $\mathrm{Bi}_2\mathrm{Se}_3$ source was first heated to $700$~$^\circ$C for $1$~min and subsequently ramped to $740$~$^\circ$C for $5$~min, while the $\mathrm{Bi}_2\mathrm{O}_3$ source was simultaneously heated to $690$~$^\circ$C and maintained for $5$~min \autocite{Liu2020}.

The $\mathrm{Bi}_2\mathrm{O}_2\mathrm{Se}$ Josephson junction devices were fabricated using electron-beam (e-beam) lithography (Goldenscope Tech Pharos 310). In situ argon milling was used to prepare the deposition surface. Under $5$~nm sticking layers of Ti (for $\mathrm{Bi}_2\mathrm{O}_2\mathrm{Se}$) were deposited in an ultrahigh-vacuum e-beam evaporator followed by $35$~nm to $45$~nm of Al, which was then covered in situ with a top layer of $15$~nm to $20$~nm of Au to form the electrodes. For devices with Cu-doped Al, we expanded the range of the e-beam movement during the Al evaporation to co-evaporate some Cu from the crucible, and the Al-to-Cu content ratio in the electrode is evaluated to be around 4 using energy-dispersive X-ray spectroscopy (EDX).

The Bernal bilayer graphene Josephson junction devices were fabricated using a dry-transfer technique with a poly(bisphenol A carbonate) (PPC)/polydimethylsiloxane (PDMS) stamp mounted on a glass slide. Hexagonal boron nitride (hBN) flakes ($10$--$40$~nm), monolayer $\mathrm{WSe_2}$ (commercial source, HQ Graphene), BBG, and graphite flakes ($3$--$10$~nm) were mechanically exfoliated onto Si/SiO$_2$ substrates. The stacking sequence from top to bottom consists of: top hBN dielectric, monolayer $\mathrm{WSe_2}$, BBG, and bottom hBN dielectric. The layers were sequentially picked up from top to bottom using the PPC film at an elevated temperature, with careful approach and slow pickup to ensure clean interfaces. The complete stack was then released onto a clean Si/SiO$_2$ substrate, and the PPC residue was removed by annealing in an Ar/H$_2$ (9:1) atmosphere at $300$~$^\circ$C for $12$~hours to eliminate residual polymer and interfacial bubbles. The Josephson junction geometry was defined by e-beam lithography, followed by reactive ion etching using a $\mathrm{CF_4/O_2}$ plasma to etch the stack down to the bottom hBN layer. Ohmic edge contacts (Ti/Cu-doped Al/Au, $5$~nm/$35$~nm/$15$~nm) were deposited by e-beam evaporation. To complete the device, a graphite/hBN stack was picked up using PPC and subsequently released onto the prefabricated Josephson junction, forming a dual-graphite-gate structure.

Transport measurements were conducted in a dilution refrigerator (Oxford Proteox) with a base temperature of $8.5$~mK equipped with a $9/1/1$~T vector magnet. We apply an AC excitation superimposed on a DC bias current and monitor the resulting differential voltage following established techniques \autocite{Hart2014}. DC sources were Keithley~2400 and Keithley~2460 source-measure units. AC excitation and measurements were performed using SR830 and NF~5645 lock-in amplifiers. In our supercurrent interference measurements, the positive switching current (top half of each SIP) was obtained from sweeping the bias current in the positive direction, while the negative switching current (bottom half of each SIP) was obtained from sweeping the bias current in the negative direction.
\par

\textbf{Simulation of the supercurrent interference pattern}

In the main text, we model the supercurrent interference pattern with constant SGFs in the leads. Considering the current entering and exiting the junction, we now extend the model to include linear profile of SGFs and show the emergence of lifted nodes (Fig. 2a). Starting again from Eqn. \eqref{lead_derivative}, we derive in the asymmetric configuration the supercurrent profiles in the leads $ I_{1,2}(x) = \pm I_{\mathrm{bias}} \left( \frac{x}{W} - \frac{1}{2} \right) $. Therefore the phase difference profile becomes

\begin{equation} 
\phi_{1,2}(x) - \phi_{1,2}(-\frac{W}{2}) = \pm L_k I_{\mathrm{bias}}\int_{-W/2}^{x} \left( \frac{a}{W} - \frac{1}{2} \right) \mathrm{d}a = 
\pm L_k I_{\mathrm{bias}} \left( \frac{x^2}{2W} - \frac{x}{2} -\frac{3W}{8} \right) .
\end{equation} 
We can absorb the constant terms into the global phase variable and arrive at

\begin{equation} 
\gamma(x) = \gamma_0 + ( k_B + L_k I_{\mathrm{bias}}) x - \frac{L_k I_{\mathrm{bias}} }{W} x^2.
\end{equation}
The resultant Josephson critical current

\begin{equation} 
I_c =  \max_{\gamma_0} \int_{-W/2}^{W/2} J_c \sin \left(\gamma_0 + ( k_B + L_k I_{\mathrm{bias}}) x - \frac{L_k I_{\mathrm{bias}} }{W} x^2 \right) \mathrm{d}x, 
\end{equation}
can be simplified to

\begin{equation} 
I_c = J_c \left| \int_{-W/2}^{W/2} e^{i ( k_B + L_k I_{\mathrm{bias}}) x - i \frac{L_k I_{\mathrm{bias}} }{W} x^2 } \mathrm{d}x \right|,
\end{equation}
which we implemented in our numerical calculation to generate the Fresnel diffraction pattern in Fig. 2c.

\clearpage

\noindent{\large\textbf{Acknowledgements}}\\
We thank the Fundamental and Interdisciplinary Disciplines Breakthrough Plan of the Ministry of Education of China (JYB2025XDXM410). This work is broadly supported by the National Natural Science Foundation of China; in particular, D. Geng acknowledges grant no. 524B2011. K.W. and T.T. acknowledge support from the JSPS KAKENHI (grants 21H05233 and 23H02052) and World Premier International Research Center Initiative (WPI), MEXT, Japan for hBN crystal fabrication/characterization.
\par\quad\par

\noindent{\large\textbf{Author contributions}}\\
H. Ren conceived the project and constructed the theory. H. Ye synthesized and fabricated the $\mathrm{Bi}_2\mathrm{O}_2\mathrm{Se}$ devices. W. He, Y. Zhang, and S. Li fabricated the graphene devices. H. Ye, W. He, K. Fan, and H. Ren conducted the transport measurements. K.W. and T.T. grew the hBN crystals. H. Ye, W. He, and H. Ren analysed the data and interpreted the results. H. Ren, Y. Pan, D. Geng, F. Yang supervised the study. H. Ye, W. He, and H. Ren prepared the manuscript with input from all authors. 
\par\quad\par

\noindent{\large\textbf{Competing interests}}\\
The authors declare no competing interests.

\newpage

\section*{Extended Data}

\renewcommand{\thefigure}{S\arabic{figure}}
\setcounter{figure}{0}

\begin{figure*}[t]
\centering
\includegraphics[width=0.9\linewidth]{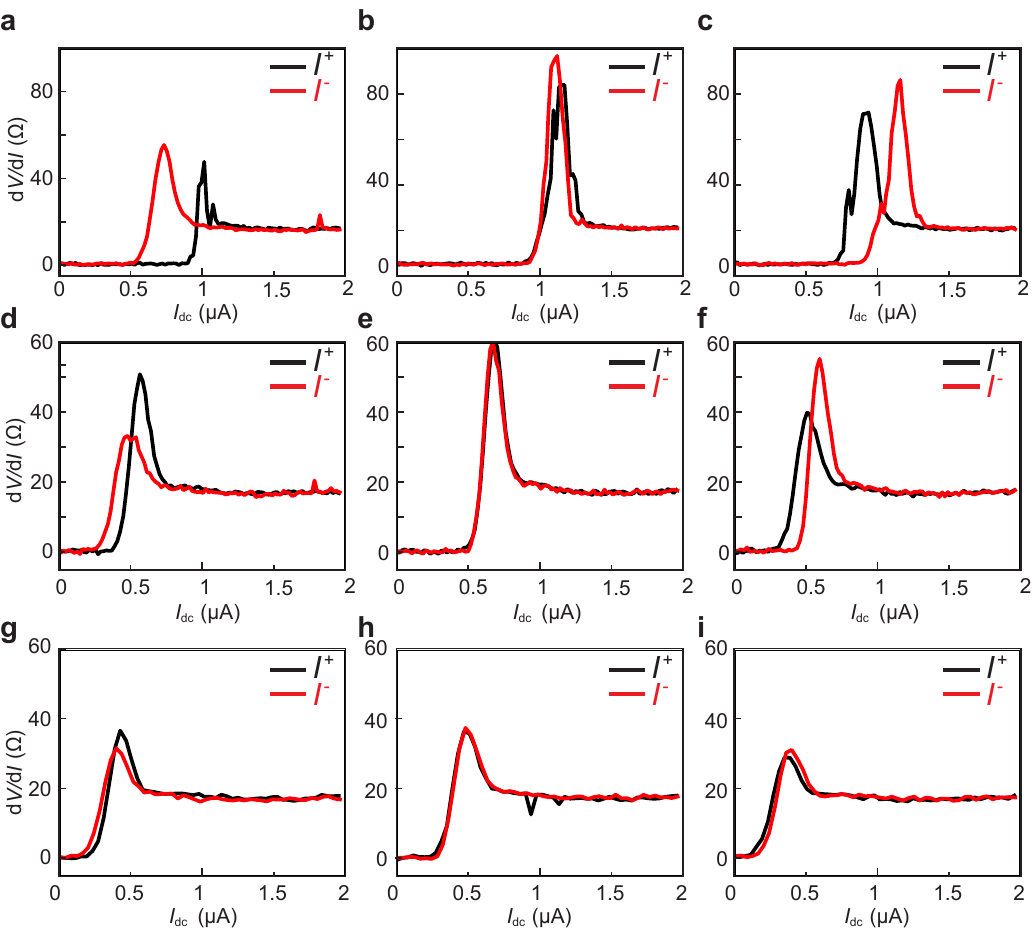} 
\caption{\textbf{In-plane magnetic field $B_x$ dependence of the JDE in JJ1 (field along current direction).} 
\textbf{a-c}, Differential resistance vs. $I_{\mathrm{dc}}$ for JJ1, using the same current configuration as in Fig.~1c, measured at $B_x = 500$~G for $B_z = -1.2$, $0$, and $+1.2$~G, respectively.
\textbf{d-f}, Differential resistance vs. $I_{\mathrm{dc}}$ at $B_x = 700$~G for $B_z = -1$, $0$, and $+1$~G, respectively. The JDE weakens as a higher $B_x$ decreases the critical currents.
\textbf{g-i}, Differential resistance vs. $I_{\mathrm{dc}}$ at $B_x = 850$~G for $B_z = -1$, $0$, and $+1$~G, respectively. Under this magnetic field, the JDE nearly vanishes.}
\label{fig_s1}
\end{figure*}
\clearpage

\begin{figure*}[t]
\centering
\includegraphics[width=0.9\linewidth]{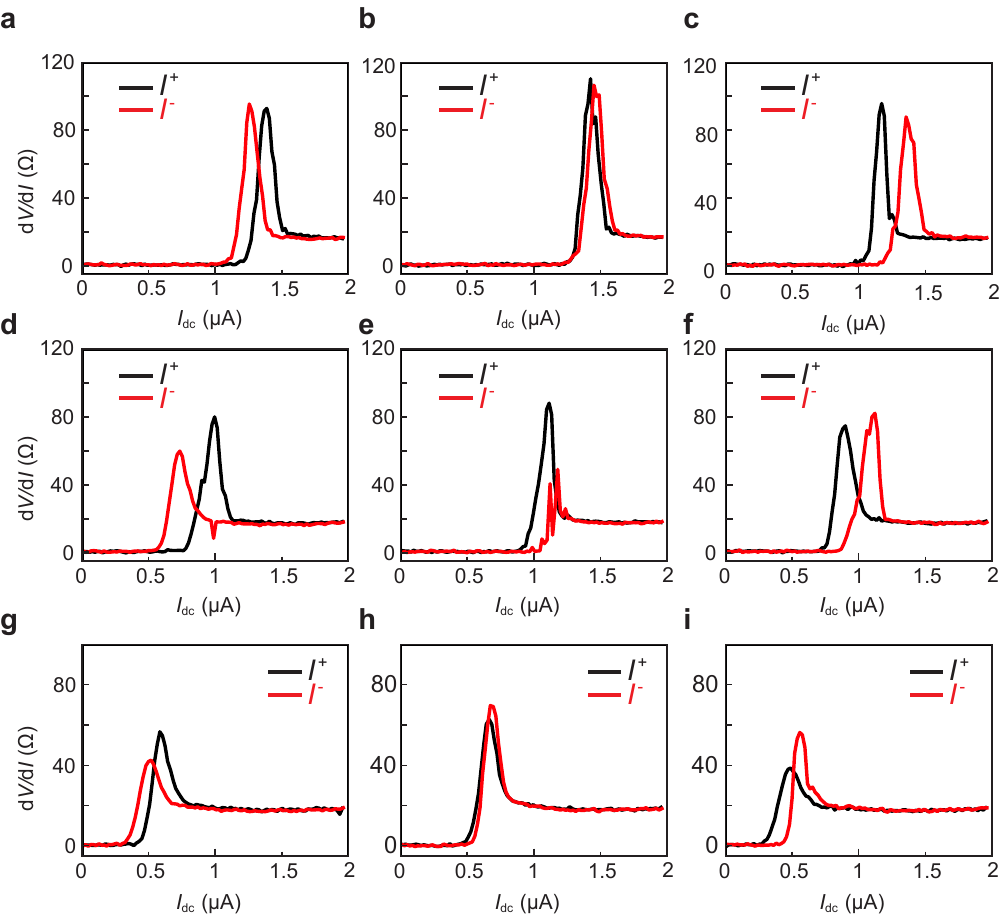} 
\caption{\textbf{In-plane magnetic field $B_y$ dependence of the JDE in JJ1 (field perpendicular to current direction).} 
\textbf{a-c}, Differential resistance vs. $I_{\mathrm{dc}}$ for JJ1, using the same current configuration as in Fig.~1c, measured at $B_y = 400$~G for $B_z = -1$, $0$, and $+1$~G, respectively.
\textbf{d-f}, Differential resistance vs. $I_{\mathrm{dc}}$ at $B_y = 1000$~G for $B_z = -1$, $0$, and $+1$~G, respectively.
\textbf{g-i}, Differential resistance vs. $I_{\mathrm{dc}}$ at $B_y = 1400$~G for $B_z = -1$, $0$, and $+1$~G, respectively. The JDE exhibits minor variation with $B_y$, and a pronounced JDE persists even at $B_y = 1400$~G.}
\label{fig_s2}
\end{figure*}
\clearpage

\begin{figure*}[t]
\centering
\includegraphics[width=0.9\linewidth]{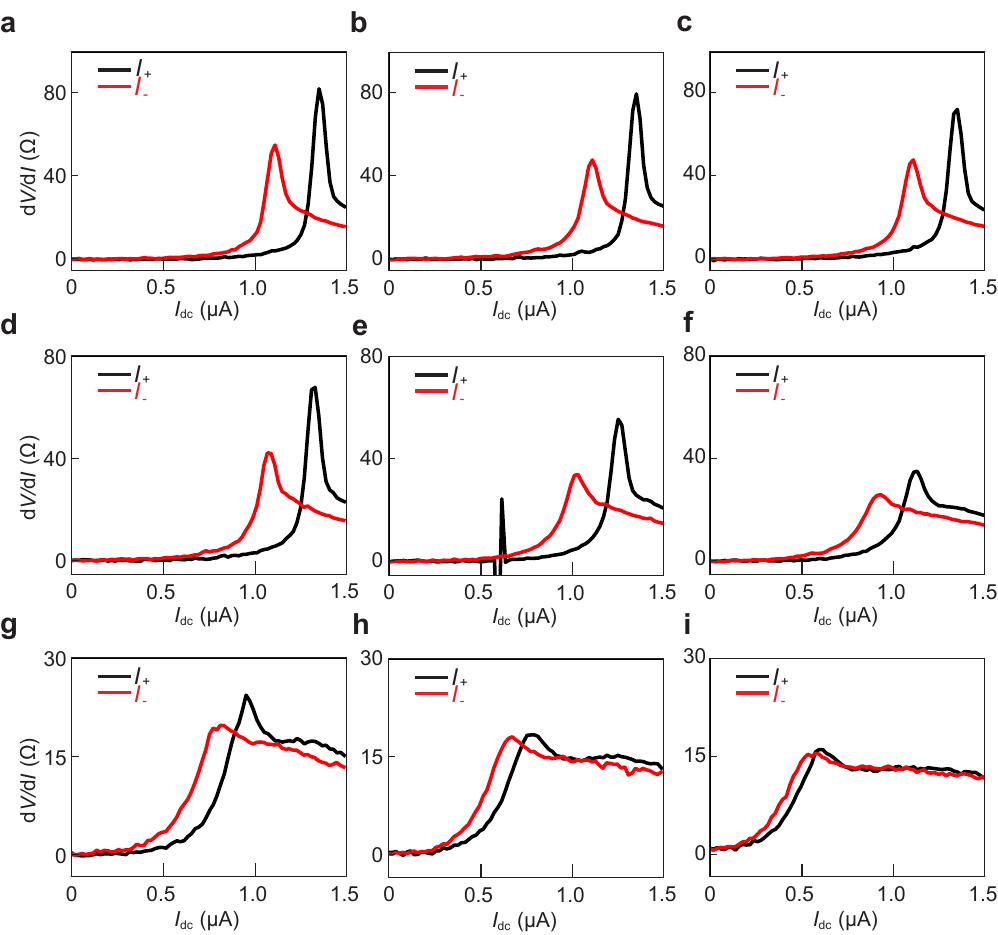} 
\caption{\textbf{Temperature dependence of JDE for device JJ3.} 
Differential resistance vs. $I_{\mathrm{dc}}$ at temperatures $T = 50$, $100$, $150$, $200$, $250$, $300$, $350$, $400$, and $450$~mK. The JDE weakens with increasing temperature and completely vanishes at $450$~mK. JJ3 is a separate Josephson junction with a conventional electrode geometry similar to that of JJ1.}
\label{fig_s3}
\end{figure*}
\clearpage

\begin{figure*}[t]
\centering
\includegraphics[width=0.6\linewidth]{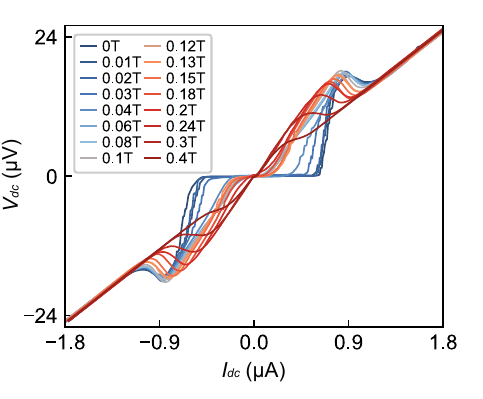} 
\caption{\textbf{$B_y$ (perpendicular to current direction) dependence in the BBG device.} Current–voltage ($I$–$V$) characteristics recorded under in-plane magnetic field $B_y$ at $V_{\text{tg}} = 0$~V and $V_{\text{bg}} = 2$~V. These data complement the $B_x$-dependent measurements shown in Fig. 3g, exhibiting a comparable suppression of the negative differential resistance (NDR) region as $B_y$ increases. The extracted NDR from this $B_y$ sweep, together with the $B_x$ data, is plotted in Fig.~3h.}
\label{fig_s4}
\end{figure*}
\clearpage

\begin{figure*}[!t]
\centering
\includegraphics[width=0.9\linewidth]{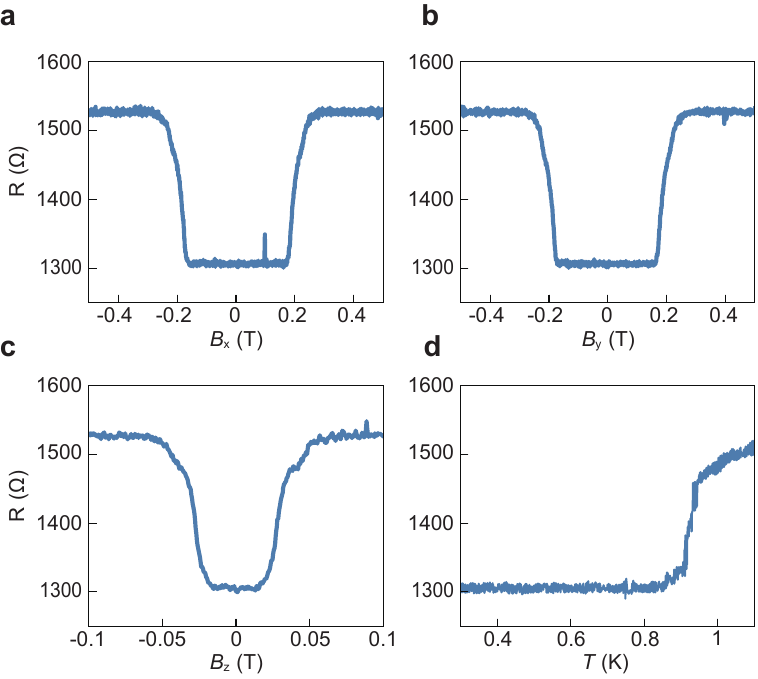} 
\caption{\textbf{Critical magnetic fields and critical temperature of the meandering electrodes presented in JJ2.} 
All measurements are performed in a two-terminal configuration, yielding a superconducting state base resistance of approximately $1300\ \Omega$, which originates primarily from non-superconducting lead resistances in the measurement setup.
\textbf{a-c}, Critical field $B_{\mathrm{c}}$ measured with the field oriented along the $x$, $y$, and $z$ directions, respectively: $B_{\mathrm{c},x}=0.187\ \mathrm{T}$, $B_{\mathrm{c},y}=0.194\ \mathrm{T}$, and $B_{\mathrm{c},z}=0.028\ \mathrm{T}$.
\textbf{d}, Resistance versus temperature curve, revealing a superconducting transition temperature $T_{\mathrm{c}} \approx 950\ \mathrm{mK}$. This value is substantially lower than the $T_{\mathrm{c}}$ of pure Al ($\sim 1.2\ \mathrm{K}$), which we attribute to the reverse proximity effect from the Au layer in the Al/Au bilayer.}
\label{fig_s5}
\end{figure*}
\clearpage

\begin{figure*}[!t]
\centering
\includegraphics[width=0.9\linewidth]{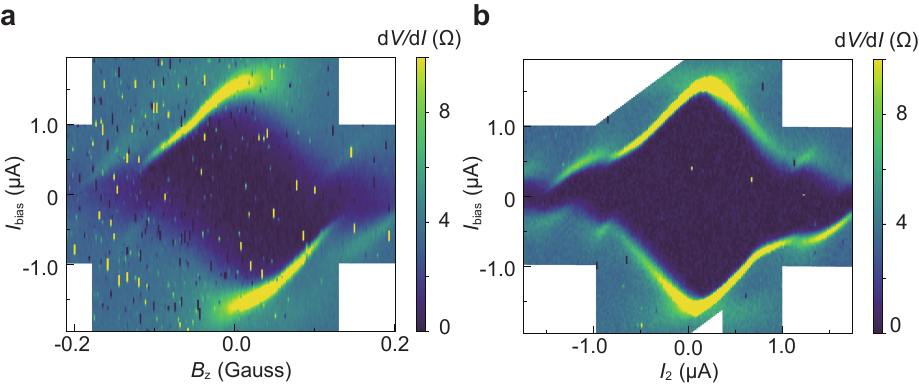} 
\caption{\textbf{Critical currents in JJ2 under symmetric current configuration.} 
\textbf{a}, SIP under an external magnetic field ($I_{\text{bias}} = I_{3} - I_{1}$, $I_{1} = I_{2} = 0$). Under this current configuration, the SGFs from the two electrodes mutually cancel each other, resulting in negligible JDE. This result demonstrates that the existence of JDE can be controlled by the current configuration.
\textbf{b}, Zero-field SIP obtained with two-terminal control ($I_{1} = -I_{2}$) and bias $I_{\text{bias}} = I_{3} - I_{1}$. This demonstrates that the SGF and magnetic flux exert identical phase modulation effects in Josephson junctions under symmetric current configuration.}
\label{fig_s6}
\end{figure*}
\clearpage

\end{spacing}

\end{document}